\documentclass[journal]{IEEEtran}

\ifCLASSINFOpdf
\else
\fi

\usepackage{amsthm} 
\usepackage{wrapfig}
\usepackage{amssymb}
\usepackage{latexsym}
\usepackage{bm, bbm} 
\usepackage{epsfig}
\usepackage{graphicx}
\usepackage{amsmath}
\usepackage{epsfig}
\usepackage{multicol}
\usepackage{psfrag}
\usepackage{float}
\usepackage{cite}
\usepackage{subfigure}
\usepackage{color}

\renewcommand{\mathbf}[1]{{\bm{#1}}}

\newcommand{\ignore}[1]{}

\newcommand{\dmins}{d_{\mathrm{min}}}
\newcommand{\dfree}{d_{\mathrm{free}}}
\newcommand{\matr}[1]{\mathbf{#1}}
\newcommand{\vect}[1]{\mathbf{#1}}
\newcommand{\code}[1]{\mathcal{#1}}
\newcommand{\cC}{\mathcal{C}}
\newcommand{\set}[1]{\mathcal{#1}}
\newcommand{\graph}[1]{\mathsf{#1}}

\newcommand{\GF}[1]{\mathbb{F}_{#1}}

\newcommand{\girth}{\operatorname{girth}}
\newcommand{\matrH}{\matr{H}}

\newcommand{\matrHconv}{\matr{H}_{\mathrm{conv}}}
\newcommand{\shortmatrHconv}{\matr{H}}
\newcommand{\codeCconv}{\code{C}_{\mathrm{conv}}}

\newcommand{\tr}{\mathsf{T}}

\newcommand{\codeCQC}[1]{\code{C}_{\mathrm{QC}}^{(r)}}

\def\wt{\operatorname{wt}}

\newcommand{\defeq}{\triangleq}

\newcommand{\vc}{\vect{c}}
\newcommand{\vcconv}{\vect{c}}

\newcommand{\vtc}{\vect{\tilde c}}
\newcommand{\vs}{\vect{s}}

\newcommand{\Z}{\mathbb{Z}}

\renewcommand{\leq}{\leqslant}
\renewcommand{\geq}{\geqslant}

\newcommand{\Zr}{\Z / r \Z}

\newcommand{\confer}{\emph{cf.}}
\newcommand{\eg}{\emph{e.g.}}

\newcommand{\etc}{\emph{etc.}}
\newcommand{\ie}{\emph{i.e.}}

\newtheorem{lemma}{Lemma}
\newtheorem{theorem}[lemma]{Theorem}

\newtheorem{corollary}[lemma]{Corollary}

\theoremstyle{plain}

\newtheorem{PreDefinition}[lemma]{{\textbf{Definition}}}
  \newenvironment{definition}%
    {\begin{PreDefinition}}{\hfill$\square$\end{PreDefinition}}

\theoremstyle{plain}

\newtheorem{PreRemark}[lemma]{{\textbf{Remark}}}
    {\begin{PreRemark}\upshape}{\hfill$\square$\end{PreRemark}}

\newtheorem{PreExample}[lemma]{{\textbf{Example}}}
  \newenvironment{example}%
    {\begin{PreExample}\upshape}{\hfill$\square$\end{PreExample}}

\newcommand{\wH}{w_{\mathrm{H}}}

\newcommand{\matzero}{{\mathbf 0}} 
\newcommand{\Ftwo}{\mathbb{F}_2} 
\newcommand{\Ftwoxmodr}{\Ftwo[x] / \langle x^r{-}1 \rangle}
\newcommand{\shortFtwoxmodr}{\Ftwo^{\langle r \rangle}[x]}
\newcommand{\Ftwoylauser}{\Ftwo(\!(y)\!)}

\newcommand{\minstarnoarg}{\operatorname{min}^{*}}
\newcommand{\minstar}[1]{\underset{#1}{\operatorname{min}^{*}}}

\newcommand{\perm}{\operatorname{perm}}

\newcommand{\neighborhood}{\partial}

\newcommand{\cover}[1]{\widetilde{#1}}

\newcommand{\cgraph}[1]{\cover{\graph{#1}}}

\newcommand{\vertsize}{J}
\newcommand{\horsize}{I}

\newcounter{mytempeqcounter}

\begin{document}

\title{Quasi-Cyclic LDPC Codes: \\
       Influence of Proto- and Tanner-Graph Structure on 
       Minimum Hamming Distance Upper Bounds}

\author{Roxana Smarandache,~\IEEEmembership{Member,~IEEE,}
        and Pascal O.~Vontobel,~\IEEEmembership{Member,~IEEE}%
  \thanks{Submitted to IEEE Transactions on Information Theory,
          January 23, 2009. Revised July 3, 2011, and August 20, 2011.
    The first author was partially supported by NSF Grants DMS-0708033 and
    TF-0830608. The second author was partially supported by NSF Grant
    CCF-0514801. The material in this paper has been presented in part at the
    2004 International Symposium on Information Theory, Chicago, IL, USA,
    June/July 2004.}%
  \thanks{R.~Smarandache is with the Department of Mathematics and Statistics,
    San Diego State University, San Diego, CA 92182, USA.  (e-mail:
    rsmarand@sciences.sdsu.edu).}%
  \thanks{P.~O.~Vontobel is with Hewlett-Packard Laboratories, 1501 Page Mill
    Road, Palo Alto, CA 94304, USA.  (e-mail: pascal.vontobel@ieee.org).}%
}

\markboth{Submitted to IEEE Transactions on Information Theory}%
         {Smarandache and Vontobel: XXXXX Title}

\maketitle

\begin{abstract}
  Quasi-cyclic (QC) low-density parity-check (LDPC) codes are an important
  instance of proto-graph-based LDPC codes. In this paper we present upper
  bounds on the minimum Hamming distance of QC LDPC codes and study how these
  upper bounds depend on graph structure parameters (like variable degrees,
  check node degrees, girth) of the Tanner graph and of the underlying
  proto-graph. Moreover, for several classes of proto-graphs we present
  explicit QC LDPC code constructions that achieve (or come close to) the
  respective minimum Hamming distance upper bounds.

  Because of the tight algebraic connection between QC codes and convolutional
  codes, we can state similar results for the free Hamming distance of
  convolutional codes. In fact, some QC code statements are established by
  first proving the corresponding convolutional code statements and then using
  a result by Tanner that says that the minimum Hamming distance of a QC code
  is upper bounded by the free Hamming distance of the convolutional code that
  is obtained by ``unwrapping'' the QC code.
\end{abstract}


\IEEEpeerreviewmaketitle

\section{Introduction}
\label{sec:introduction:1}

\IEEEPARstart{Q}{uasi-cyclic} (QC) low-density parity-check (LDPC) codes
represent an important class of codes within the family of LDPC
codes~\cite{Gallager:63}. The first graph-based code construction that yielded
QC codes was presented by Tanner in~\cite{Tanner:99:2}; although that code
construction was presented in the context of repeat-accumulate codes, it was
easy to generalize the underlying idea to LDPC codes in order to obtain QC
LDPC codes~\cite{Fan:00:1, MacKay:Davey:01:1, Fossorier:04:1,
  Milenkovic:Prakash:Vasic:03:1}. The simplicity with which QC LDPC codes can
be described makes them attractive for implementation and analysis purposes.

A QC LDPC code of length $n = \horsize r$ can be described by a $\vertsize r
\times \horsize r$ (scalar) parity-check matrix that is formed by a $\vertsize
\times \horsize$ array of $r \times r$ circulant matrices. Clearly, by
choosing these circulant matrices to be low-density, the parity-check matrix
will also be low-density.

With the help of the well-known isomorphism between the ring of circulant
matrices over some field $\mathbb{F}$ and the ring of $\mathbb{F}$-polynomials
modulo $x^r - 1$ (see, \eg, \cite{Lally:Fitzpatrick:01:1}), a QC LDPC code can
equally well be described by a polynomial parity-check matrix of size
$\vertsize \times \horsize$. In the remainder of the paper we will mainly work
with the polynomial parity-check matrix of a QC LDPC code and not with the
(scalar) parity-check matrix. Another relevant concept in this paper will be
the weight matrix associated with a polynomial parity-check matrix; this
weight matrix is a $\vertsize \times \horsize$ integer matrix whose entries
indicate the number of terms of the corresponding polynomial in the polynomial
parity-check matrix.

Early papers on QC LDPC codes focused mainly on polynomial parity-check
matrices whose weight matrix contained only ones. Such polynomial parity-check
matrices are known as monomial parity-check matrices because all entries are
monomials, \ie, polynomials with exactly one term. For this class of QC LDPC
codes it was soon established that the minimum Hamming distance is always
upper bounded by $(\vertsize \! + \! 1)!$~\cite{MacKay:Davey:01:1,
  Fossorier:04:1, Smarandache:Vontobel:04:1}.

In this paper we study polynomial parity-check matrices with more general
weight matrices by allowing the entries of the weight matrix to be $0$, $1$,
$2$, or $3$ (and sometimes larger). This is equivalent to allowing the entries
of the polynomial parity-check matrix to be the zero polynomial, to be a
monomial, to be a binomial, or to be a trinomial (and sometimes a polynomial
with more nonzero coefficients). The main theme will be to analyze the minimum
Hamming distance of such codes, in particular by studying upper bounds on the
minimum Hamming distance and to see how these upper bounds depend on other
code parameters like the girth of the Tanner graph. We will obtain upper
bounds that are functions of the polynomial parity-check matrix and upper
bounds that are functions of the weight matrix. The latter results are in
general weaker but they give good insights into the dependency of the minimum
Hamming distance on the structure of the weight matrix. For example, for
$\vertsize = 3$ we show that there are weight matrices that are different from
the all-one weight matrix (but with the same column and row sum) that yield
minimum Hamming distance upper bounds that are larger than the above-mentioned
$(\vertsize \! + \! 1)!$ bound. By constructing some codes that achieve this
upper bound we are able to show that the discrepancies in upper bounds are not
spurious.

Being able to obtain minimum Hamming distance bounds as a function of the
weight matrix is also interesting because the weight matrix is tightly
connected to the concept of proto-graphs and LDPC codes derived from
them~\cite{Thorpe:03:1, Thorpe:Andrews:Dolinar:04:1}. Proto-graph-based code
constructions start with a proto-graph that is described by a $\vertsize
\times \horsize$ incidence matrix whose entries are non-negative integers and
where a ``$0$'' entry corresponds to no edge, a ``$1$'' entry corresponds to a
single edge, a ``$2$'' entry corresponds to two parallel edges, \etc. (Such an
incidence matrix is also known as a proto-matrix.) Once such a proto-graph is
specified, a proto-graph-based LDPC code is then defined to be the code whose
Tanner graph~\cite{Tanner:81} is some $r$-fold graph cover~\cite{Massey:77:1,
  Stark:Terras:96:1} of that proto-graph.

It is clear that the construction of QC LDPC codes can then be seen as a
special case of the proto-graph-based construction: first, the weight matrix
corresponds to the proto-matrix, \ie, the incidence matrix of the proto-graph;
secondly, the $r$-fold cover is obtained by restricting the edge permutations
to be cyclic.

A main reason for the attractiveness of QC LDPC codes is that they can be
encoded efficiently using approaches like in~\cite{Li:Chen:Zeng:Lin:Fong:06:1}
and decoded efficiently using belief-propagation-based decoding
algorithms~\cite{Kschischang:Frey:Loeliger:01} or LP-based decoding
algorithms~\cite{Feldman:03:1, Feldman:Wainwright:Karger:05:1,
  Vontobel:Koetter:07:1, Taghavi:Siegel:08:1}. Although the behavior of these
decoders is mostly dominated by pseudo-codewords~\cite{Wiberg:96,
  Forney:Koetter:Kschischang:Reznik:01:1, Frey:Koetter:Vardy:01:1,
  Koetter:Vontobel:03:1, Vontobel:Koetter:05:1:subm,
  Koetter:Li:Vontobel:Walker:07:1, Smarandache:Vontobel:07:1,
  Kelley:Sridhara:07:1} and the (channel-dependent) pseudo-weight of
pseudo-codewords, the minimum Hamming distance still plays an important role
because it characterizes undetectable errors and it provides an upper bound on
the minimum pseudo-weight of a Tanner graph representing a code.

Although the main focus of this paper is on QC codes, we can state analogous
results for convolutional codes. Besides the interest that these statements
generate on their own, from a theorem proving point of view these results are
helpful because some of our results for QC codes are most easily proven by
first proving the corresponding results for convolutional codes. From a
technical point of view, this stems from the fact that convolutional codes are
defined by parity-check matrices over a field (more precisely, the field
$\Ftwoylauser$ specified in Section~\ref{sec:notation:1:basic}), whereas QC
codes are defined by parity-check matrices over rings (more precisely, the
ring $\shortFtwoxmodr$ specified in Section~\ref{sec:notation:1:basic}), and
that consequently there are more linear algebra tools available to handle
convolutional codes than to handle QC codes.

The remainder of this paper is structured as follows.\footnote{This overview
  mentions only QC code results and omits the analogous convolutional code
  results.} Section~\ref{sec:definition:1} introduces important concepts and
the notation that will be used throughout the paper. Thereafter,
Section~\ref{sec:minimum:hamming:distance:upper:bounds:1} presents the two
main results of this paper. Both results are upper bounds on the minimum
Hamming distance of a QC code: whereas in the case of Theorem~\ref{th:bound:1}
the upper bound is a function of the polynomial parity-check matrix of the QC
code, in the case of Theorem~\ref{th:bound:2} the upper bound is a function of
the weight matrix of the QC code only. The following two sections are then
devoted to the study of special cases of these results. Namely,
Section~\ref{sec:type:1:qc:ldpc:codes:1} focuses on so-called type-$1$ QC LDPC
codes (\ie, QC LDPC codes where the weight matrix entries are at most $1$) and
Section~\ref{sec:type:2:and:type:3:qc:ldpc:codes:1} focuses on so-called
type-$2$ and type-$3$ QC LDPC codes (\ie, QC LDPC codes where the weight
matrix entries are at most $2$ and $3$, respectively). We will show how we can
obtain type-$2$ and type-$3$ codes from type-$1$ codes having the same
regularity and possibly better minimum Hamming distance properties.
Section~\ref{sec:effect:small:cycles:minimum:distance} investigates the
influence of cycles on minimum Hamming distance bounds. Finally,
Section~\ref{sec:type:1:qc:ldpc:codes:based:on:double:cover:1} discusses a
promising construction of type-$1$ QC LDPC codes based on type-$2$ or type-$3$
QC LDPC codes. In fact, we suggest a sequence of constructions starting with a
type-$1$ code that exhibits good girth and minimum Hamming distance
properties, or that has good performance under message-passing iterative
decoding. We construct a type-$2$ or type-$3$ code with the same regularity
and higher Hamming distance upper bound, and from this we obtain a new
type-$1$ code with possibly larger minimum Hamming
distance. Section~\ref{sec:conclusions:1} concludes the paper.  The appendix
contains the longer proofs and also one section
(\confer~Appendix~\ref{sec:graph:covers:1}) that lists some results with
respect to graph covers.

\section{Definitions}
\label{sec:definition:1}

This section formally introduces the objects that were discussed in
Section~\ref{sec:introduction:1}, along with some other definitions that will
be used throughout the paper.

\subsection{Sets, Rings, Fields, Vectors, and Matrices}
\label{sec:notation:1:basic}

We use the following sets, rings, and fields: for any positive integer $L$,
$[L]$ denotes the set $\{ 0, 1, \ldots, L {-}1 \}$; $\Z$ is the ring of
integers; for any positive integer $r$, $\Zr$ is the ring of integers modulo
$r$; $\Ftwo$ is the Galois field of size $2$; $\Ftwo[x]$ is the ring of
polynomials with coefficients in $\Ftwo$ and indeterminate $x$; $\Ftwoxmodr$
is the ring of polynomials in $\Ftwo[x]$ modulo $x^r - 1$, where $r$ is a
positive integer; and $\Ftwoylauser$ is the field of formal Laurent series
over $\Ftwo$, \ie, the set $\bigl\{ \sum_{\ell=d}^{\infty} a_{\ell} y^{\ell}
\bigm| d \in \Z, \ a_{\ell} \in \Ftwo, \ell \geq d \bigr\}$ with the usual
rules for addition and multiplication. We will often use the notational
short-hand $\shortFtwoxmodr$ for $\Ftwoxmodr$.

By $\Ftwo^n$ and $\Ftwo^{m \times n}$ we will mean, respectively, a row vector
over $\Ftwo$ of length $n$ and a matrix over $\Ftwo$ of size $m \times n$,
with a similar meaning given to $\shortFtwoxmodr^n$, $\shortFtwoxmodr^{m
  \times n}$, $\Ftwoylauser^n$, and $\Ftwoylauser^{m \times n}$. In the
following we will use the convention that indices of vector entries start at
$0$ (and not at $1$), with a similar convention for row and column indices of
matrix entries.

For any matrix $\matr{M}$, we let $\matr{M}_{\set{R}, \set{S}}$ be the
sub-matrix of $\matr{M}$ that contains only the rows of $\matr{M}$ whose index
appears in the set $\set{R}$ and only the columns of $\matr{M}$ whose index
appears in the set $\set{S}$. If $\set{R}$ equals the set of all row indices
of $\matr{M}$, we will omit in $\matr{M}_{\set{R}, \set{S}}$ the set $\set{R}$
and we will simply write $\matr{M}_{\set{S}}$.  Moreover, we will use the
short-hand $\matr{M}_{\set{S} \setminus i}$ for $\matr{M}_{\set{S} \setminus
  \{ i \}}$.

As usual, the $\min$ operator gives back the minimum value of a list of
values.\footnote{If the list is empty then $\min$ gives back $+\infty$.} In
the following, we will also use a more specialized minimum operator, namely
the $\minstarnoarg$ operator that gives back the minimum value of all nonzero
entries in a list of values.\footnote{If the list is empty or if zero is the
  only value appearing in the list then $\minstarnoarg$ gives back
  $+\infty$. In particular, for lists containing only non-negative values, as
  will be the case in the remainder of this paper, the $\minstarnoarg$
  operator gives back the smallest positive entry of the list if the list
  contains positive entries, otherwise it gives back $+\infty$.}

\subsection{Weights}
\label{sec:notation:1:weights}

The weight $\wt\bigl( c(x) \bigr) \in \Z$ of a polynomial $c(x) \in \Ftwo[x]$
equals the number of nonzero coefficients of $c(x)$. Similarly, the weight
$\wt\bigl( c(x) \bigr) \in \Z$ of a polynomial $c(x) \in \shortFtwoxmodr$
equals the weight $\wt\bigl( c'(x) \bigr)$ of the (unique) minimal-degree
polynomial $c'(x) \in \Ftwo[x]$ that fulfills $c'(x) = c(x) \ (\text{in
  $\shortFtwoxmodr$})$.

Let $\vc(x) = \bigl( c_0(x), c_1(x), \ldots, c_{\horsize-1}(x) \bigr) \in
\shortFtwoxmodr^{\horsize}$ be a length-$\horsize$ polynomial vector. Then the
weight vector $\wt\bigl( \vc(x) \bigr) \in \Z^\horsize$ of $\vc(x)$ is a
length-$\horsize$ vector with the $i$-th entry equal to $\wt\bigl( c_i(x)
\bigr)$. Similarly, let $\matrH(x) = \bigl[ h_{j,i}(x) \bigr]_{j,i} \in
\shortFtwoxmodr^{\vertsize \times \horsize}$ be a size-$\vertsize {\times}
\horsize$ polynomial matrix. Then the weight matrix $\wt\bigl( \matrH(x)
\bigr) \in \Z^{\vertsize \times \horsize}$ of $\matrH(x)$ is a $\vertsize
{\times} \horsize$-matrix with the entry in row $j$ and column $i$ equal to
$\wt\bigl[ h_{j,i}(x) \bigr]$.

The Hamming weight $\wH(\vc)$ of a vector $\vc$ is the number of nonzero
entries of $\vc$. In the case of a polynomial vector $\vc(x) = \bigl( c_0(x),
c_1(x), \ldots, c_{\horsize-1}(x) \bigr) \in \Ftwo[x]^{\horsize}$, the Hamming
weight $\wH\bigl( \vc(x) \bigr)$ is defined to be the sum of the weights of
its polynomial entries, \ie, $\wH\bigl( \vc(x) \bigr) =
\sum_{i=0}^{\horsize-1} \bigl(\wt ( \vc(x) ) \bigr)_i =
\sum_{i=0}^{\horsize-1} \wt\bigl( c_i(x) \bigr)$.

Analogous definitions are used for the weight of an element of $\Ftwoylauser$,
the weight of vectors over $\Ftwoylauser$, \etc.

\subsection{QC Codes}
\label{sec:notation:1:codes}

All codes in this paper will be binary linear codes. As usual, a block code
$\code{C}$ of length $n$ can be specified through a (scalar) parity-check
matrix $\matrH \in \GF{2}^{m \times n}$, \ie, $\code{C} = \bigl\{ \vc \in
\Ftwo^n \ \bigl| \ \matrH \cdot \vc^\tr = \vect{0}^\tr \bigr. \bigr\}$, where
${}^\tr$ denotes transposition. This code has rate at least $1 - \frac{m}{n}$
and its minimum Hamming distance (which equals the minimum Hamming weight
since the code is linear) will be denoted by $\dmins(\code{C})$.

Let $\vertsize$, $\horsize$, and $r$ be positive integers. Let $\code{C}$ be a
code of length $\horsize r$ that possesses a parity-check matrix $\matrH$ of
the form
\begin{align*}
  \matrH
    &= \begin{bmatrix}
         \matrH_{0,0}   & \matrH_{0,1}  & \cdots & \matrH_{0,\horsize-1} \\
         \matrH_{1,0}   & \matrH_{1,1}  & \cdots & \matrH_{1,\horsize-1} \\
         \vdots          & \vdots          & \ddots & \vdots \\
         \matrH_{\vertsize-1,0} & \matrH_{\vertsize-1,1} & \cdots & 
           \matrH_{\vertsize-1,\horsize-1}
       \end{bmatrix}
        \in \Ftwo^{\vertsize r \times \horsize r},
\end{align*}
where the sub-matrices $\matrH_{j,i} \in \Ftwo^{r \times r}$ are circulant.
Such a code is called quasi-cyclic (QC) because applying circular shifts to
length-$r$ sub-blocks of a codeword gives a codeword again.  Because
$\matrH_{j,i}$ is circulant, it can be written as the sum $\matrH_{j,i} =
\sum_{s=0}^{r-1} h_{j,i,s,0} \cdot \matr{I}_s$, where $h_{j,i,s,0}$ is the
entry of $\matrH_{j,i}$ in row $s$ and column $0$, and where $\matr{I}_s$ is
the $s$ times cyclically left-shifted identity matrix of size $r \times r$.

With a parity-check matrix $\matrH \in \Ftwo^{\vertsize r \times \horsize r}$
of a QC code we associate the polynomial parity-check matrix $\matrH(x) \in
\shortFtwoxmodr^{\vertsize \times \horsize}$
\begin{align*}
  \matrH(x)
    &= \begin{bmatrix}
         h_{0,0}(x)   & h_{0,1}(x)  & \cdots & h_{0,\horsize-1}(x) \\
         h_{1,0}(x)   & h_{1,1}(x)  & \cdots & h_{1,\horsize-1}(x) \\
         \vdots      & \vdots      & \ddots & \vdots \\
         h_{\vertsize-1,0}(x) & h_{\vertsize-1,1}(x) & \cdots & 
           h_{\vertsize-1,\horsize-1}(x)
       \end{bmatrix},
\end{align*}
where $h_{j,i}(x) \defeq \sum_{s=0}^{r-1} h_{j,i,s,0} x^s$. Moreover, with any
vector $\vc = (c_{0,0}, \ldots, c_{0,r-1}, \ldots, c_{\horsize-1,0}, \ldots,
c_{\horsize-1,r-1}) \in \Ftwo^{\horsize r}$ we associate the polynomial vector
\begin{align*}
  \vc(x)
    &= \big(
         c_0(x), c_1(x), \ldots, c_{\horsize-1}(x)
       \big)
       \in \shortFtwoxmodr^n,
\end{align*}
where $c_i(x) \defeq \sum_{s=0}^{r-1} c_{i,s} x^s$. It can easily be
checked that the condition
\begin{align*}
  \matrH \cdot \vc^\tr
    &= \vect{0}^\tr
         \quad \text{(in $\Ftwo$)}
\end{align*}
is equivalent to the condition
\begin{align*}
  \matrH(x) \cdot \vc(x)^\tr
    &= \vect{0}^\tr
         \quad \text{(in $\shortFtwoxmodr$)},
\end{align*}
giving us an alternate way to check if a (polynomial) vector is a codeword.

The following classification was first introduced
in~\cite{Smarandache:Vontobel:04:1}.

\begin{definition}
  \label{def:type:QC:code:1}

  Let $M$ be some positive integer. We say that a polynomial parity-check
  matrix $\matrH(x)$ of a QC LDPC code is of type $M$ if all the entries of
  the associated weight matrix $\wt\bigl( \matrH(x) \bigr)$ are at most
  $M$. Moreover, we say that a QC LDPC code is of type $M$ if it is defined by
  a polynomial parity-check matrix of type $M$.
\end{definition}

Equivalently, $\matrH(x)$ is of type $M$ if for each polynomial entry in
$\matrH(x)$ the number of nonzero coefficients is at most $M$. In particular,
the polynomial parity-check matrix $\matrH(x)$ is of type $1$
(in~\cite{Smarandache:Vontobel:04:1} we also called them ``type I'') if
$\matrH(x)$ contains only the zero polynomial and monomials. Moreover, the
polynomial parity-check matrix $\matrH(x)$ is of type $2$
(in~\cite{Smarandache:Vontobel:04:1} we also called them ``type II'') if
$\matrH(x)$ contains only the zero polynomial, monomials, and binomials. If
$\matrH(x)$ contains only monomials then it will be called a monomial
parity-check matrix. (Obviously, a monomial parity-check matrix is a type-$1$
polynomial parity-check matrix.)

\subsection{Convolutional Codes}
\label{sec:notation:1:convolutional:codes}

A convolutional code $\codeCconv$ can be described by a polynomial
parity-check matrix $\matrHconv(y) \in \Ftwoylauser^{\vertsize \times
  \horsize}$; the codewords of $\codeCconv$ are then the polynomial vectors
$\vcconv(y) \in \Ftwoylauser^I$ that satisfy\footnote{Although ``formal
  Laurent series parity-check matrix'' and ``formal Laurent series vector''
  would be more precise, we use ``polynomial parity-check matrix'' and
  ``polynomial vector'' also in the context of convolutional codes.}
\begin{align*}
  \matrHconv(y) \cdot \vc(y)^\tr
    &= \vect{0}^\tr
         \quad \text{(in $\Ftwoylauser$)}.
\end{align*}
The free Hamming distance of $\codeCconv$ will be denoted by
$\dfree(\codeCconv)$. Moreover, a convolutional code whose (polynomial)
parity-check matrix is sparse will be called a convolutional LDPC code and we
extend the classification of polynomial parity-check matrices in
Definition~\ref{def:type:QC:code:1} from QC codes to convolutional codes.

The main interest of the present paper in convolutional codes is the fact that
QC codes can be ``unwrapped'' to yield convolutional codes~\cite{Tanner:87:1}
(see also~\cite{Levy:Costello:93:1,
  Esmaeili:Gulliver:Secord:Mahmoud:98:1}). In mathematical terms,
``unwrapping'' means to associate with a QC code $\code{C}$ defined by some
polynomial parity-check matrix $\matrH(x) \in \shortFtwoxmodr^{\vertsize
  \times \horsize}$ the convolutional code $\codeCconv$ defined by the
parity-check matrix $\matrHconv(y) \in \Ftwoylauser^{\vertsize \times
  \horsize}$, where
\begin{align*}
  \matrHconv(y)
    &\defeq
       \matrH(x)|_{x = y}.
\end{align*}
In other words, $\matrHconv(y)$ is obtained by replacing all appearances of
$x$ (and its powers) in $\matrH(x)$ by $y$ (and its powers). Note that the
weight matrices of $\matrH(x)$ and $\matrHconv(y)$ are the same, \ie,
$\wt\bigl( \matrH(x) \bigr) = \wt\bigl( \matrHconv(y) \bigr)$.\footnote{Here
  and in the following we assume that $\matrH(x)$ is given in a form where the
  exponents that appear in $\matrH(x)$ are at least $0$ and strictly smaller
  than $r$.}

A theorem by Tanner~\cite{Tanner:87:1} allows one then to relate the minimum
Hamming distance of the QC code $\code{C}$ to the free Hamming distance of the
above-defined convolutional code $\codeCconv$, namely
\begin{align}
  \dmins(\code{C})
    &\leq
       \dfree(\codeCconv).
         \label{eq:Tanner:dmin:dfree:1}
\end{align}
(See~\cite{Smarandache:Pusane:Vontobel:Costello:09:1} for the usage of this
theorem in the context of QC LDPC and convolutional LDPC codes, along with
generalizations of it to different notions of minimum pseudo-weights.) 

There is a simple algebraic reason why in the present paper we are interested
in the above-mentioned connection between QC codes and convolutional
codes. Namely, since the entries of $\matrHconv(y)$ are from some field,
notions like linear independence and rank are well defined for this matrix. In
particular, the zero-ness/nonzero-ness of determinants of square sub-matrices
of $\matrHconv(y)$ allow us to reach conclusions about the linear
dependence/independence of the rows and columns of these sub-matrices. Such
conclusions can in general not be reached for the sub-matrices of $\matrH(x)$,
which is a matrix with entries in some commutative ring (in particular, a ring
with zero divisors).

\subsection{Graphs}
\label{sec:notation:1:graphs}

With a parity-check matrix $\matrH$ we associate a Tanner
graph~\cite{Tanner:81} in the usual way: for every code bit we draw a variable
node, for every parity-check we draw a check node, and we connect a variable
node and a check node by an edge if and only if the corresponding entry in
$\matrH$ is nonzero. Similarly, the Tanner graph associated with a polynomial
parity-check matrix $\matrH(x)$ is simply the Tanner graph associated with the
corresponding (scalar) parity-check matrix $\matrH$.

As usual, the degree of a vertex is the number of edges incident to it and an
LDPC code is called $(d_1, d_2)$-regular if all variable nodes have
degree~$d_1$ and all check nodes have degree~$d_2$. Otherwise we will say that
the code is irregular. Moreover, a simple cycle of a graph will be a
backtrackless, tailless, closed walk in the graph, and the length of such a
cycle is defined to be equal to the number of visited vertices (or,
equivalently, the number of visited edges). The girth of a graph is then the
length of the shortest simple cycle of the graph.

The above-mentioned concepts are made more concrete with the help of the
following example.

\begin{figure}
  \begin{center}
    \epsfig{file=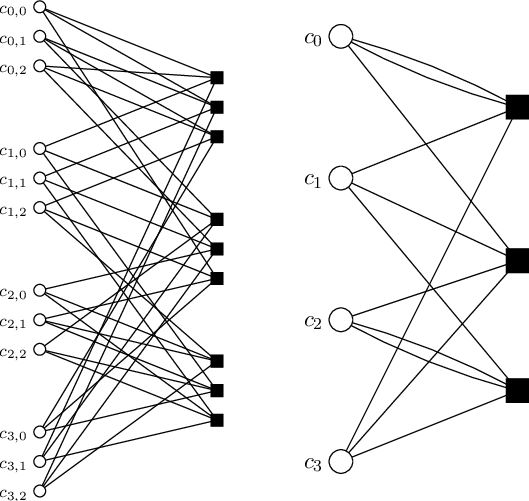, height=6cm}
  \end{center}
  \caption{Left: Tanner graph of a length-$12$ QC LDPC code. It is a triple
    cover of the proto-graph shown on the right. Right: Proto-graph of the
    Tanner graph shown on the left.}
  \label{fig:qc:ldpc:code:and:proto:graph:1}
\end{figure}

\begin{example}
  \label{ex:introductory:example:1}

  Let $\code{C}$ be a length-$12$ QC code that is described by the
  parity-check matrix
  \begin{align*}
    \matrH
      &\defeq
         \left[
           \begin{array}{ccc|ccc|ccc|ccc}
             1 & 0 & 1 &  1 & 0 & 0 &  0 & 0 & 0 &  0 & 1 & 0 \\ 
             1 & 1 & 0 &  0 & 1 & 0 &  0 & 0 & 0 &  0 & 0 & 1 \\ 
             0 & 1 & 1 &  0 & 0 & 1 &  0 & 0 & 0 &  1 & 0 & 0 \\ 
           \hline
             0 & 1 & 0 &  1 & 0 & 0 &  0 & 0 & 1 &  0 & 1 & 0 \\ 
             0 & 0 & 1 &  0 & 1 & 0 &  1 & 0 & 0 &  0 & 0 & 1 \\ 
             1 & 0 & 0 &  0 & 0 & 1 &  0 & 1 & 0 &  1 & 0 & 0 \\ 
           \hline
             0 & 0 & 0 &  0 & 0 & 1 &  1 & 1 & 0 &  0 & 0 & 1 \\ 
             0 & 0 & 0 &  1 & 0 & 0 &  0 & 1 & 1 &  1 & 0 & 0 \\ 
             0 & 0 & 0 &  0 & 1 & 0 &  1 & 0 & 1 &  0 & 1 & 0
           \end{array}
         \right].
  \end{align*}
  Clearly, $\vertsize = 3$, $\horsize = 4$, and $r = 3$ for this code and so
  $\matrH$ can also be written like
  \begin{align*}
    \matrH
      &= \begin{bmatrix}
           \matr{I}_0 + \matr{I}_1 & \matr{I}_0 & \matr{0} & \matr{I}_2 \\
           \matr{I}_2 & \matr{I}_0 & \matr{I}_1 & \matr{I}_2 \\
           \matr{0}   & \matr{I}_1 & \matr{I}_0 + \matr{I}_2 & \matr{I}_1
         \end{bmatrix},
  \end{align*}
  where $\matr{I}_s$, $s = 0, 1, \ldots, r{-}1$, are $s$-times cyclically
  left-shifted $r \times r$ identity matrices. The corresponding polynomial
  parity-check matrix is
  \begin{align*}
    \matrH(x)
      &= \begin{bmatrix}
           x^0 + x^1 & x^0 & 0 & x^2 \\
           x^2 & x^0 & x^1 & x^2 \\
           0   & x^1 & x^0 + x^2 & x^1
         \end{bmatrix},
  \end{align*}
  and the weight matrix is
  \begin{align*}
    \wt\big( \matrH(x) \big)
      &= \begin{bmatrix}
           2 & 1 & 0 & 1 \\
           1 & 1 & 1 & 1 \\
           0 & 1 & 2 & 1
         \end{bmatrix}.
  \end{align*}
  The Tanner graph associated with $\matrH$ or $\matrH(x)$ is shown in
  Figure~\ref{fig:qc:ldpc:code:and:proto:graph:1}~(left). We observe that all
  variable nodes have degree~$3$ and all check nodes have degree~$4$,
  therefore $\code{C}$ is a $(3,4)$-regular LDPC code. (Equivalently, all
  columns of $\matrH$ have weight $3$ and all rows of $\matrH$ have weight
  $4$.)
\end{example}

The proto-graph associated with a polynomial parity-check matrix $\matrH(x)$
is a graphical representation of the weight matrix $\wt\bigl( \matrH(x)
\bigr)$ in the following way. It is a graph where for each column of
$\matrH(x)$ we draw a variable node, for each row of $\matrH(x)$ we draw a
check node, and the number of edges between a variable node and a check node
equals the corresponding entry in $\wt\bigl( \matrH(x) \bigr)$.

\begin{example}
  \label{ex:introductory:example:2}

  Continuing Example~\ref{ex:introductory:example:1}, the proto-graph of
  $\matrH(x)$ is shown in
  Figure~\ref{fig:qc:ldpc:code:and:proto:graph:1}~(right). Clearly, the weight
  matrix $\wt\big( \matrH(x) \big)$ is the incidence matrix of this latter
  graph. We observe that all variable nodes have degree~$3$ and all check
  nodes have degree~$4$. (Equivalently, all column sums (in $\Z$) of $\wt\big(
  \matrH(x) \big)$ equal $3$ and all row sums (in $\Z$) of $\wt\big(
  \matrH(x) \big)$ equal $4$.)
\end{example}

An important concept for this paper is that of the so-called graph covers, see
the next definition.

\begin{definition}[See, \eg, \cite{Massey:77:1, Stark:Terras:96:1}] 

  Let $\graph{G}$ be a graph with vertex set $\set{V}(\graph{G})$ and edge set
  $\set{E}(\graph{G})$, and let $\neighborhood(v)$ denote the set of adjacent
  vertices of a vertex $v \in \set{V}(\graph{G})$. An {\em unramified, finite
    cover}, or, simply, a {\em cover} of a (base) graph $\graph{G}$ is a graph
  $\cgraph{G}$ along with a surjective map $\phi: \cgraph{G} \to \graph{G}$,
  which is a graph homomorphism, \ie, which takes adjacent vertices of
  $\cgraph{G}$ to adjacent vertices of $\graph{G}$ such that, for each vertex
  $v \in \set{V}(\graph{G})$ and each $\cover{v} \in \phi^{-1}(v)$, the
  neighborhood $\neighborhood(\cover{v})$ of $\cover{v}$ is mapped bijectively
  to $\neighborhood(v)$. For a positive integer $r$, an {\em $r$-cover} of
  $\graph{G}$ is an unramified finite cover $\phi: \cgraph{G} \to \graph{G}$
  such that, for each vertex $v \in \set{V}(\graph{G})$ of $\graph{G}$,
  $\phi^{-1}(v)$ contains exactly $r$ vertices of $\cgraph{G}$. An $r$-cover
  of $\graph{G}$ is sometimes also called an $r$-sheeted covering of
  $\graph{G}$ or a cover of $\graph{G}$ of degree $r$.\footnote{It is
    important not to confuse the degree of a covering and the degree of a
    vertex.}
\end{definition}

\begin{example}
  \label{ex:introductory:example:3}

  Continuing Examples~\ref{ex:introductory:example:1}
  and~\ref{ex:introductory:example:2}, we note that the graph in
  Figure~\ref{fig:qc:ldpc:code:and:proto:graph:1}~(left) is a $3$-cover of the
  graph in Figure~\ref{fig:qc:ldpc:code:and:proto:graph:1}~(right). Therefore,
  the code $\code{C}$ is a proto-graph-based code. It can easily be checked
  visually that all edge permutations that were used to define this $3$-cover
  are cyclic permutations, confirming that the code $\code{C}$ is indeed
  quasi-cyclic.
\end{example}

Tanner graphs can also be defined for convolutional codes (see,
\eg,~\cite{Pusane:Smarandache:Vontobel:Costello:11:1}); in particular, the
paper~\cite{Pusane:Smarandache:Vontobel:Costello:11:1} discusses some
connections between the Tanner graph of a QC code and the Tanner graph of a
convolutional code that is obtained by ``unwrapping'' the QC code.

We conclude this subsection by emphasizing that graph covers have been used in
two different ways in the context of LDPC codes: on the one hand, they have
been used for constructing LDPC codes (like in this paper), on the other hand
they have been used to analyze message-passing iterative decoders (like
in~\cite{Koetter:Vontobel:03:1, Vontobel:Koetter:05:1:subm}).

\subsection{Determinants and Permanents}
\label{sec:notation:1:determinants:and:permanents}

The determinant of an $m \times m$-matrix $\matr{B} = [b_{j,i}]_{j,i}$
over some commutative ring is defined to be
\begin{align*}
  \det(\matr{B})
    &= \sum_{\sigma}
         \operatorname{sgn}(\sigma)
         \prod_{j \in [m]}
           b_{j,\sigma(j)},
\end{align*}
where the summation is over all $m!$ permutations of the set $[m]$, and where
$\operatorname{sgn}(\sigma)$ equals $+1$ if $\sigma$ is an even permutation
and equals $-1$ if $\sigma$ is an odd permutation.

The permanent of an $m \times m$-matrix $\matr{B} = [b_{j,i}]_{j,i}$ over some
commutative ring is defined to be
\begin{align*}
  \perm(\matr{B})
    &= \sum_{\sigma}
         \prod_{j \in [m]}
         b_{j,\sigma(j)},
\end{align*}
where the summation is over all $m!$ permutations of the set $[m]$.

Clearly, for any matrix $\matr{B}$ with elements from a commutative ring of
characteristic $2$ it holds that $\det(\matr{B}) = \perm(\matr{B})$.

\section{Minimum Hamming Distance Upper Bounds}
\label{sec:minimum:hamming:distance:upper:bounds:1}

This section contains the two main theoretical results of this paper, namely
Theorems~\ref{th:bound:1} and~\ref{th:bound:2}. More precisely, given some QC
code with polynomial parity-check matrix $\matrH(x)$ and minimum Hamming
distance $\dmins(\code{C})$, Theorem~\ref{th:bound:1} presents an upper bound
on $\dmins(\code{C})$ as a function of the entries of $\matrH(x)$ and
Theorem~\ref{th:bound:2} presents an upper bound on $\dmins(\code{C})$ as a
function of the entries of $\wt\bigl( \matrH(x) \bigr)$. The upper bound of
Theorem~\ref{th:bound:2} is in general weaker than the upper bound of
Theorem~\ref{th:bound:1}, however, it is interesting to see that the weight
matrix alone can already give nontrivial bounds on the achievable minimum
Hamming distance. These theorems also present analogous results for the free
Hamming distance of convolutional codes.

In Sections~\ref{sec:type:1:qc:ldpc:codes:1}
and~\ref{sec:type:2:and:type:3:qc:ldpc:codes:1}, we will discuss the
implications of these two theorems on codes with type-$1$, type-$2$, and
type-$3$ polynomial parity-check matrices. Moreover, in
Section~\ref{sec:effect:small:cycles:minimum:distance} we will show how the
upper bounds in Theorems~\ref{th:bound:1} and~\ref{th:bound:2} can be
strengthened by taking some graph structure information (like cycles) into
account.

We start with a simple technique to construct codewords of codes described by
polynomial parity-check matrices; this extends a codeword construction
technique by MacKay and Davey~\cite[Theorem~2]{MacKay:Davey:01:1}. (Note that
the paper~\cite{MacKay:Davey:01:1} deals with codes that are described by
scalar parity-check matrices composed of commuting permutation sub-matrices,
of which parity-check matrices composed of cyclically shifted identity
matrices are a special case. However, and as we show in this paper, their
techniques can be suitably extended to codes that are described by scalar
parity-check matrices composed of \emph{any} circulant matrices, and therefore
to codes that are described by polynomial parity-check matrices.)

\begin{lemma}
  \label{lemma:codeword:construction:1}

  Let $\code{C}$ be the QC code defined by the polynomial parity-check matrix
  $\matrH(x) \in \shortFtwoxmodr^{\vertsize \times \horsize}$. Let $\set{S}$
  be an arbitrary size-$(\vertsize {+}1)$ subset of $[\horsize]$ and let
  $\vect{c}(x) = \big( c_0(x), c_1(x), \ldots, c_{\horsize-1}(x) \big) \in
  \shortFtwoxmodr^\horsize$ be a length-$\horsize$ vector defined
  by\footnote{Because the ring $\shortFtwoxmodr$ has characteristic $2$, we
    could equally well define $c_i(x) \defeq \det\big( \matrH_{\set{S}
      \setminus i}(x) \big)$ if $i \in \set{S}$.}
  \begin{align*}
    c_i(x)
      &\defeq
         \begin{cases} 
           \perm\big( \matrH_{\set{S} \setminus i}(x) \big) 
             & \text{if $i \in \set{S}$} \\
           0                                     
             & \text{otherwise}
         \end{cases}.
  \end{align*}
  Then $\vect{c}(x)$ is a codeword in $\code{C}$.

  An analogous construction yields codewords of the convolutional code
  $\codeCconv$ defined by the polynomial parity-check matrix $\matrHconv(y) \in
  \Ftwoylauser^{\vertsize \times \horsize}$.
\end{lemma}

\begin{IEEEproof}
  Let $\set{S} = \{ i_0, i_1, \ldots, i_\vertsize \}$ be the chosen
  size-$(\vertsize {+}1)$ subset. In order to verify that $\vect{c}(x)$ is a
  codeword in $\code{C}$, we need to show that the syndrome $\vs^\tr(x) =
  \matrH(x) \cdot \vc^\tr(x)$ (in $\shortFtwoxmodr$) is the all-zero
  vector. For any $j \in [\vertsize]$, we can express the $j$-th component of
  $\vect{s}(x)$ as follows
  \begin{align*}
    s_j(x) 
       = \sum_{i \in [\horsize]}
           h_{j,i}(x) c_i(x)
      &= \sum_{i \in \set{S}}
           h_{j,i}(x)
           \cdot
           \perm\big( \matrH_{\set{S} \setminus i}(x) \big) \\
      &= \sum_{i \in \set{S}}
           h_{j,i}(x)
           \cdot
           \det\big( \matrH_{\set{S} \setminus i}(x) \big),
  \end{align*}
  where in the last step we used the fact that for commutative rings with
  characteristic $2$ the permanent equals the determinant. Observing that
  $s_j(x)$ is the co-factor expansion of the determinant of the $|\set{S}|
  \times |\set{S}|$-matrix
  \begin{align*}
    \left[
    \begin{array}{cccc}
      h_{j,i_0}(x)   & h_{j,i_1}(x)   & \cdots & h_{j,i_\vertsize }(x) \\ 
      \hline
      h_{0,i_0}(x)   & h_{0,i_1}(x)   & \cdots & h_{0,i_\vertsize }(x) \\ 
      h_{1,i_0}(x)   & h_{1,i_1}(x)   & \cdots & h_{1,i_\vertsize }(x) \\
      \vdots        & \vdots        & \cdots & \vdots \\ 
      h_{\vertsize-1,i_0}(x) & h_{\vertsize-1,i_1}(x) & \cdots & 
        h_{\vertsize-1,i_\vertsize }(x) \\ 
    \end{array}
    \right],  
  \end{align*}
  and noting that this latter matrix is singular (because at least two rows
  are equal), we obtain the result that $\vs(x) = \vect{0}$ and that $\vc(x)$
  is indeed a codeword in $\code{C}$, as promised.

  Because $\Ftwoylauser$ is a field, and therefore a commutative ring, the
  same argument holds also for a code like $\codeCconv$ that is defined by
  a parity-check matrix over $\Ftwoylauser$.
\end{IEEEproof}

\mbox{}

With the help of the codeword construction technique in
Lemma~\ref{lemma:codeword:construction:1} we can easily obtain the bound in
Theorem~\ref{th:bound:1}: simply construct the list of all codewords
corresponding to all size-$(\vertsize {+}1)$ subsets $\set{S}$ of
$[\horsize]$, and use the fact that the minimum Hamming distance of
$\code{C}$~/ the free Hamming distance of $\codeCconv$ is upper bounded by the
minimum Hamming weight of all \emph{nonzero} codewords in this list.

\begin{theorem}
  \label{th:bound:1}

  Let $\code{C}$ be the QC code defined by the polynomial
  parity-check matrix $\matrH(x) \in \shortFtwoxmodr^{\vertsize \times
    \horsize}$. Then the minimum Hamming distance of $\code{C}$ is upper
  bounded as follows
  \begin{align}
    \dmins(\code{C})
      &\leq
         \minstar{\set{S} \subseteq [\horsize] 
                    \atop |\set{S}| = \vertsize +1} \ 
           \sum_{i \in \set{S}}
             \wt
               \Big(
                 \perm\big( \matrH_{\set{S} \setminus i}(x) \big)
               \Big).
                 \label{eq:bound:1} 
  \end{align}

  Let $\codeCconv$ be the convolutional code defined by the polynomial
  parity-check matrix $\matrHconv(y) \in \Ftwoylauser^{\vertsize \times
    \horsize}$. Then the free Hamming distance of $\codeCconv$ is upper
  bounded as follows
  \begin{align}
    \hskip-0.20cm
    \dfree(\codeCconv)
      &\leq
         \minstar{\set{S} \subseteq [\horsize]
                    \atop |\set{S}| = \vertsize +1} \ 
           \sum_{i \in \set{S}}
             \wt
               \Big(
                 \perm\big( (\matrHconv)_{\set{S} \setminus i}(y) \big)
               \Big).
                 \label{eq:bound:1:convolutional:code} 
  \end{align}
  (Note that for $\matrHconv(y) = \matrH(x) |_{x = y}$ the right-hand sides
  of~\eqref{eq:bound:1} and~\eqref{eq:bound:1:convolutional:code} need not be
  equal.)
\end{theorem}

\begin{IEEEproof}
  We start by proving the QC code part of this theorem. Let $\set{S}$ be a
  size-$(\vertsize \! + \! 1)$ subset of $[\horsize]$ and let $\vect{c}(x)$ be
  the corresponding codeword constructed according to
  Lemma~\ref{lemma:codeword:construction:1}. The result in the theorem
  statement follows by noting that $\vect{c}(x)$ has Hamming weight
  \begin{align*}
    \wH\big( \vc(x) \big)
      &= \sum_{i \in [\horsize]}
           \wt\big( c_i(x) \big)
       = \sum_{i \in \set{S}}
           \wt\big( c_i(x) \big) \\
      &= \sum_{i \in \set{S}}
           \wt
             \Big(
               \perm\big( \matrH_{\set{S} \setminus i}(x) \big)
             \Big).
  \end{align*}
  
  The convolutional code part of this theorem then follows from the
  observation that $\Ftwoylauser$ is a field (and therefore a commutative
  ring), and so the above derivation also holds for the parity-check matrix
  $\matrHconv(y)$.
\end{IEEEproof}

\mbox{}

Let us emphasize that it is important to have the $\minstarnoarg$ operator
in~\eqref{eq:bound:1}, and not just the $\min$ operator. The reason is that
the upper bound is based on constructing codewords of the code $\code{C}$ and
evaluating their Hamming weight. For some polynomial parity-check matrices
some of these constructed codewords may equal the all-zero codeword and
therefore have Hamming weight zero: clearly, such constructed codewords are
irrelevant for upper bounding the minimum Hamming distance and therefore must
be discarded. This is done with the help of the $\minstarnoarg$
operator. (Similar statements can be made with respect
to~\eqref{eq:bound:1:convolutional:code}.)

The next theorem, Theorem~\ref{th:bound:2}, gives a minimum/free Hamming
distance upper bound which is easier to compute than~\eqref{eq:bound:1}
and~\eqref{eq:bound:1:convolutional:code} and which depends only on the weight
matrix associated with $\matrH(x)$ and $\matrHconv(y)$, respectively. In
particular, this bound does \emph{not} depend on $r$, the size of the
circulant matrices in the scalar parity-check matrix $\matrH$ corresponding to
$\matrH(x)$. The bound says that the minimum/free Hamming distance is upper
bounded by the minimum nonzero sum of the permanents of all $\vertsize \times
\vertsize$ sub-matrices of a chosen $\vertsize \times (\vertsize {+}1)$
sub-matrix of the weight matrix, the minimum being taken over all such
possible $\vertsize \times (\vertsize {+}1)$ sub-matrices of the weight
matrix.

\begin{theorem}
  \label{th:bound:2}

  Let $\code{C}$ be a QC code with polynomial parity-check matrix $\matrH(x)
  \in \shortFtwoxmodr^{\vertsize \times \horsize}$ and let $\matr{A} \defeq
  \wt\bigl( \matrH(x) \bigr)$, or, let $\codeCconv$ be a convolutional code
  with polynomial parity-check matrix $\matrHconv(y) \in \Ftwoylauser^{\vertsize
    \times \horsize}$ and let $\matr{A} \defeq \wt\bigl( \matrHconv(y)
  \bigr)$. Then
  \begin{align}
    \left.
      \begin{array}{c}
        \dmins(\code{C}) \\[0.2cm]
        \dfree(\codeCconv)
      \end{array}
    \right\}
      &\leq
         \minstar{\set{S}
                    \subseteq [\horsize] \atop |\set{S}| = \vertsize +1} \ 
           \sum_{i \in \set{S}} \ 
             \perm
               \left(
                 \matr{A}_{\set{S} \setminus i}
               \right).
                 \label{eq:bound:2}
  \end{align}
  In particular, if $\matrHconv(y) = \matrH(x) |_{x = y}$ then
  \begin{align*}
    \dmins(\code{C})
      &\leq
         \dfree(\codeCconv)
       \leq
         \minstar{\set{S}
                    \subseteq [\horsize] \atop |\set{S}| = \vertsize +1} \ 
           \sum_{i \in \set{S}} \ 
             \perm
               \left(
                 \matr{A}_{\set{S} \setminus i}
               \right).
  \end{align*}
\end{theorem}

\begin{IEEEproof}
  See Appendix~\ref{sec:proof:th:bound:2}.
\end{IEEEproof}

\mbox{}

Again, as in Theorem~\ref{th:bound:1}, it is important to have the
$\minstarnoarg$ operator in Theorem~\ref{th:bound:2} and not just the $\min$
operator. This time the reasoning is a bit more involved, though, and we refer
the reader to the proof of Theorem~\ref{th:bound:2} for details.\footnote{We
  are grateful to O.~Y.~Takeshita for pointing out to us that in earlier (and
  also less general) versions of Theorem~\ref{th:bound:1} and
  Theorem~\ref{th:bound:2} (\confer~\cite{Smarandache:Vontobel:04:1}) the
  $\min$ operator has to be replaced by the $\minstarnoarg$ operator, see
  also~\cite{Takeshita:05:1:subm}.}

Note that the upper bound in~\eqref{eq:bound:1} depends on $r$ (because the
computations are done modulo $x^r - 1$), whereas the bound
in~\eqref{eq:bound:2} \emph{does not} depend on $r$.

Usually, the expressions in~\eqref{eq:bound:1}
and~\eqref{eq:bound:1:convolutional:code} yield upper bounds that are not
larger than the upper bounds from~\eqref{eq:bound:2}. However, this does not
need to happen. For example, there are polynomial parity-check matrices for
which~\eqref{eq:bound:1} and~\eqref{eq:bound:1:convolutional:code} evaluate to
$+\infty$, whereas~\eqref{eq:bound:2} evaluates to some finite number.

Based on Theorems~\ref{th:bound:1} and~\ref{th:bound:2}, the following recipe
can be formulated for the construction of QC LDPC codes with good minimum
Hamming distance. (A similar recipe can be given for the construction of
convolutional LDPC codes with good free Hamming distance.)
\begin{itemize}

\item Search for a suitable weight matrix with the help of
  Theorem~\ref{th:bound:2}.

\item Among all polynomial parity-check matrices with this weight matrix, find
  a suitable polynomial parity-check matrix with the help of
  Theorem~\ref{th:bound:1}.

\item Verify explicitly if the minimum Hamming distance of the code of the
  found polynomial parity-check matrix really equals (or comes close to) the
  minimum Hamming distance promised by the upper bound in
  Theorem~\ref{th:bound:1}.

\end{itemize}

This recipe is especially helpful in the case where one is searching among
type-$M$ polynomial parity-check matrices with small $M$, say $M \in \{ 1, 2,
3 \}$. In such cases it is to be expected that there is not much difference in
the upper bounds~\eqref{eq:bound:1} and~\eqref{eq:bound:2}.  For type-$M$
polynomial parity-check matrices with larger $M$, however, we do not expect
that the upper bounds~\eqref{eq:bound:1} and~\eqref{eq:bound:2} are close. The
reason is that when computing $\perm\bigl( \matrH_{\set{S} \setminus i}(x)
\bigr)$ in~\eqref{eq:bound:1} there will be many terms that cancel each
other. Anyway, when constructing QC LDPC codes, type-$M$ polynomial
parity-check matrices with large $M$ are somewhat undesirable because of the
relatively small girth of the corresponding Tanner graph. In particular, it is
well known that the Tanner graph of a polynomial parity-check matrix whose
weight matrix contains at least one entry of weight $3$ (or larger) has girth
at most $6$ (see also Theorem~\ref{th:cycles:patterns:1}).

\section{Type-I QC/Convolutional Codes}
\label{sec:type:1:qc:ldpc:codes:1}

In this section we specialize the results of the previous section to the case
of type-$1$ parity-check matrices.

\begin{corollary}
  \label{cor:type:I:minimum:distance:upper:bound:1}

  Let $\code{C}$ be a type-$1$ QC code with polynomial parity-check matrix
  $\matrH(x) \in \shortFtwoxmodr^{\vertsize \times \horsize}$ and let
  $\matr{A} \defeq \wt\bigl( \matrH(x) \bigr)$, or, let $\codeCconv$ be a
  type-$1$ convolutional code with polynomial parity-check matrix $\matrHconv(y)
  \in \Ftwoylauser^{\vertsize \times \horsize}$ and let $\matr{A} \defeq
  \wt\bigl( \matrHconv(y) \bigr)$. Then
  \begin{align}
    \left.
      \begin{array}{c}
        \dmins(\code{C}) \\[0.2cm]
        \dfree(\codeCconv)
      \end{array}
    \right\}
      &\leq
         (\vertsize +1) !.
    \label{eq:bound:type:1:1}
  \end{align}
\end{corollary}

\begin{IEEEproof}
  See Appendix~\ref{sec:proof:cor:type:I:minimum:distance:upper:bound:1}.
\end{IEEEproof}

\mbox{}

The rest of this section will be devoted to QC codes; however, analogous
results can also be stated for convolutional codes.

Let us evaluate the minimum Hamming distance upper bounds that we have
obtained so far for some type-$1$ QC code polynomial parity-check
matrices. (Actually, the following polynomial parity-check matrices happen to
be \emph{monomial} parity-check matrices, \ie, polynomial parity-check
matrices where all entries of the corresponding weight matrices equal~$1$.)

\begin{example}
  \label{ex:bound:1}

  Let $r\geq 9$. Consider the $(2, 4)$-regular length-$4r$ QC code $\code{C}$
  given by the polynomial parity-check matrix
  \begin{align*}
   {\matrH}(x)
       = \begin{bmatrix}
           x      & x^2   & x^4  & x^8 \\
           x^5    & x^{6} & x^{3} & x^7
         \end{bmatrix}
  \end{align*}
  and the $(2, 4)$-regular length-$4r$ QC code $\code{C}'$ given by the
  polynomial parity-check matrix
  \begin{align*}
    \matrH'(x)
      = \begin{bmatrix}
           x      & x^2   & x^4  & x^8 \\
           x^6    & x^{5} & x^{3} & x^9
    \end{bmatrix}.
  \end{align*}
  According to~\eqref{eq:bound:1}, the minimum Hamming distance of $\code{C}$
  is upper bounded by
  \begin{align*}
    \dmins
      &\leq
         \minstarnoarg\!
           \left\{\!\!\!
             \begin{array}{l}
               \wt(x^{4}{+}x^9) + \wt(x^{5}{+}x^{10}) + \wt(x^7{+}x^7)     , \\
               \wt(x^{9}{+}x^{14}) + \wt(x^{8}{+}x^{13}) + \wt(x^7{+}x^7),\\
               \wt(x^{11}{+}x^{11}) + \wt(x^{8}{+}x^{13}) + \wt(x^{4}{+}x^9), \\
               \wt(x^{11}{+}x^{11}) + \wt(x^{9}{+}x^{14}) + \wt(x^{5}{+}x^{10})  
             \end{array}
          \!\!\!\right\} \\
      &= \minstarnoarg
           \{ 4, 4, 4,4 \}
       = 4,
  \end{align*}
  and the minimum Hamming distance of $\code{C}'$ is upper bounded by
  \begin{align*}
    \dmins
      &\leq
         \minstarnoarg\!
           \left\{\!\!\!\!
             \begin{array}{l}
               \wt(x^{5}{+}x^{9})\!+\!\wt(x^{4}{+}x^{10})\!+\!\wt(x^6{+}x^8), \\
               \wt(x^{11}{+}x^{13})\!+\!\wt(x^{10}{+}x^{14})\!+\!\wt(x^6{+}x^8), \\
               \wt(x^{13}{+}x^{11})\!+\!\wt(x^{10}{+}x^{14})\!+\!\wt(x^4{+}x^{10}), \\
               \wt(x^{13}{+}x^{11})\!+\!\wt(x^{11}{+}x^{13})\!+\!\wt(x^{5}{+}x^9)
             \end{array} 
           \!\!\!\right\}\\
      &= \minstarnoarg
           \{ 6, 6, 6, 6 \}
       = 6.
  \end{align*}
  However, in both cases the bound in~\eqref{eq:bound:2} gives
  \begin{align*}
    \dmins
      &\leq
         \big\{
           (1{+}1) + (1{+}1) + (1{+}1)
         \big\}
       = 6, 
  \end{align*}
  since both polynomial parity-check matrices have the same weight
  matrix. Similarly, in both cases the bound in~\eqref{eq:bound:type:1:1}
  gives
  \begin{align*}
    \dmins
      &\leq
         (2{+}1)!
       = 6, 
  \end{align*}
  since both polynomial parity-check matrices have $\vertsize  = 2$.

  In conclusion, we see that a $2 {\times} 4$ monomial parity-check matrix can
  yield a QC code with minimum Hamming distance at most $6$. However, when the
  entries of the polynomial parity-check matrix are not chosen suitably, as is
  the case for $\matrH(x)$, then the minimum Hamming distance upper bound
  in~\eqref{eq:bound:1} is strictly smaller than the minimum Hamming distance
  upper bound in~\eqref{eq:bound:2}.

  For completeness, we computed the minimum Hamming distance of the two
  codes,\footnote{Here and elsewhere in the paper, we compute the minimum
    distance of various QC codes with the help of suitable Magma
    programs~\cite{Bosma:Cannon:Playoust:97:1}. For analyzing the free
    distance of convolutional codes, a suitable program is, \eg,
    BEAST~\cite{Bocharova:Handlery:Johannesson:Kdryashov:04:1}.} and obtained
  $2$ for the first code (\eg, $(0, 0, x^4, 1)$ is a codeword), and $4$ for
  the second code for most values of $r$.
\end{example}

Let us discuss another example.

\begin{example}
  \label{ex:shortened:tanner:code:1} 

  Let $r \geq 26$ and let the $(3,4)$-regular QC LDPC code $\code{C}$ be given
  by the polynomial parity-check matrix $\matrH(x)\in
  \shortFtwoxmodr^{3\times 4}$
  \begin{align*}
    \matrH(x)
      &= \begin{bmatrix}
           x      & x^2    & x^4    & x^8 \\
           x^5    & x^{10} & x^{20} & x^9 \\
           x^{25} & x^{19} & x^7    & x^{14}
         \end{bmatrix}.
  \end{align*}
  (This code was obtained by shortening the last $r$ positions of the
  $(3,5)$-regular type-$1$ QC LDPC code of length $5r$ presented
  in~\cite{Tanner:Sridhara:Fuja:01:1}.\footnote{Note that
    in~\cite{Tanner:Sridhara:Fuja:01:1}, $r=31$, and the code parameters are
    $[155, 64, 20]$. Also note that by shortening a code, the girth of the
    associated Tanner graph cannot decrease.})  Evaluating the bounds
  in~\eqref{eq:bound:2} and~\eqref{eq:bound:type:1:1} for this polynomial
  parity-check matrix, we see that the minimum Hamming distance is upper
  bounded by $24$, and for suitable choices of $r$ this upper bound is indeed
  achieved. We computed the minimum Hamming distance of the code for different
  values of $r$ and obtained that $r = 31$ is the smallest such choice. The
  code obtained for $r=31$ has parameters $[124, 33, 24]$. The minimum Hamming
  distance and rate for this and some other values of $r$ are listed in the
  following table.

  \begin{center}
    \begin{tabular}{|c||@{\ }c@{\ }|@{\ }c@{\ }|@{\ }c@{\ }|@{\ }c@{\ }|@{\ }c@{\ }|@{\ }c@{\ }|}
      \hline
        $r$                 & $26$ & $27$ & $28$ & $29$ & $30$ & $31$ \\
      \hline
      \hline
        $\dmins(\code{C})$  & $18$ & $14$ & $16$ & $18$ & $8$  & $24$ \\
      \hline
        \text{rate}         & 0.269 & 0.287 & 0.268 & 0.267 & 0.283 & 0.266 \\
      \hline
    \end{tabular}
  \end{center}
\end{example}

As we have seen from the above examples, the minimum Hamming distance upper
bound~\eqref{eq:bound:1} can be strictly smaller than the upper
bound~\eqref{eq:bound:2}. However, the upper bound~\eqref{eq:bound:2} is
computed more easily, and it provides an upper bound on the Hamming distance
of all QC codes having the same weight matrix and therefore also the same
proto-graph.

Applying Corollary~\ref{cor:type:I:minimum:distance:upper:bound:1} to QC codes
with monomial parity-check matrices shows that for such codes the minimum
Hamming distance is upper bounded by $(\vertsize {+}1)!$. We note that this
result was previously presented by MacKay and Davey~\cite{MacKay:Davey:01:1}
and discussed by Fossorier~\cite{Fossorier:04:1}. However, as we show in this
paper, their techniques can be suitably extended to QC codes that are
described by scalar parity-check matrices composed of \emph{any} circulant
matrices, and therefore to codes that are described by polynomial parity-check
matrices.

\begin{example}
  \label{example:relatively:high:rate:qc:code:1}

  It is clear that the higher the rate of a code is, the more difficult it is
  to achieve the upper bound in
  Corollary~\ref{cor:type:I:minimum:distance:upper:bound:1}. However, the QC
  code defined by the polynomial parity-check matrix
  \begin{align*}
    \matrH(x)
      &\defeq
         \begin{bmatrix}
           x^{0}  & x^{19} & x^{13} & x^{20} & x^{4} & x^{15} & x^{56} \\
           x^{18} & x^{9}  & x^{0}  & x^{47} & x^{0} & x^{18} & x^{8} \\
           x^{14} & x^{0}  & x^{10} & x^{13} & x^{0} & x^{0}  & x^{7}
         \end{bmatrix}
  \end{align*}
  with $r = 111$ shows that there exist also QC codes with design rate $4/7$
  that achieve the minimum Hamming distance upper bound in
  Corollary~\ref{cor:type:I:minimum:distance:upper:bound:1}, \ie, $\dmins =
  24$. (This example is taken
  from~\cite[Table~III]{Bocharova:Hug:Johannesson:Kudryashov:Satyukov:11:1:subm}.)
\end{example}

We would like to warn the reader that we do \emph{not} claim that the
``recipe'' given at the end of
Section~\ref{sec:minimum:hamming:distance:upper:bounds:1} is an optimal
strategy for obtaining QC codes that achieve the upper bounds presented in
this paper. In particular, instead of fixing the polynomial parity-check
matrix and increasing $r$, it might be a good idea to change the polynomial
parity-check matrix as well with increasing $r$, thereby allowing the degrees
of the polynomials to grow with $r$. Such a strategy might yield codes that
achieve the upper bounds for smaller $r$; however, investigating this approach
is beyond the scope of this paper.

\section{Type-II and Type-III \\
               QC/Convolutional Codes}
\label{sec:type:2:and:type:3:qc:ldpc:codes:1}

After having discussed minimum/free Hamming distance upper bounds for type-$1$
QC/convolutional codes in the previous section, we now present similar results
for type-$2$ and type-$3$ QC/convolutional codes. In particular, we classify
all possible weight matrices of $(3,4)$-regular QC/convolutional codes with a
$3 \times 4$ polynomial parity-check matrix.

We start our investigations with the following motivating example.

\begin{example}
  \label{ex:binomial:optimal:1}

  In Example~\ref{ex:shortened:tanner:code:1} we saw that the minimum Hamming
  distance of type-$1$ $(3,4)$-regular QC codes with a $3 \times 4$ polynomial
  parity-check matrix cannot surpass $24$. In this example we show that
  type-$2$ $(3,4)$-regular QC codes with a $3 \times 4$ polynomial parity-check
  matrix can have minimum Hamming distance strictly larger than $24$.
  Namely, consider the code $\code{C}'$ with parity-check matrix
  \begin{align}
   \hskip-0.15cm
   \matrH'(x) 
      &\defeq \begin{bmatrix}
           x + x^2 & 0               & x^4             & x^8 \\
           x^5     & x^9             & x^{10} + x^{20}  & 0 \\
           0       & x^{25} + x^{19}  & 0               & x^7 + x^{14}
    \end{bmatrix}.
      \label{eq:binomial:optimal:1}
  \end{align}
  (This polynomial parity-check matrix was obtained from the parity-check
  matrix $\matrH(x)$ in Example~\ref{ex:shortened:tanner:code:1} by
  pairing some monomials into binomials and replacing with $0$ the positions
  left, careful to preserve the $(3,4)$-regularity.) The corresponding weight
  matrix is
  \begin{align*}
    \matr{A}'
      &= \begin{bmatrix}
           2 & 0 & 1 & 1 \\
           1 & 1 & 2 & 0 \\
           0 & 2 & 0 & 2
         \end{bmatrix},
  \end{align*}
  and, according to~\eqref{eq:bound:2}, yields the following minimum Hamming
  distance upper bound
  \begin{align*}
    \dmins\left( \cC' \right)
      &\leq
         \minstarnoarg
           \left\{
             10 + 6 + 10 + 6
           \right\}
       = 32.
  \end{align*} 
  For small $r$, the corresponding QC code does not attain this bound,
  however, for $r = 46$ one can verify that the resulting QC code attains the
  optimal minimum Hamming distance $\dmins = 32$. This is a $[184,47,32]$ code
  of rate $0.2554$.
\end{example}

After this introductory example, let us have a more systematic view of the
possible weight matrices of $(3,4)$-regular QC/convolutional codes and the
minimum/free Hamming distance upper bounds that they yield.

\begin{corollary}
  \label{cor:type:II:minimum:distance:upper:bound:1}

  Let $\code{C}$ be a $(3,4)$-regular type-$2$ QC code with polynomial
  parity-check matrix $\matrH(x) \in \shortFtwoxmodr^{3 \times 4}$ and let
  $\matr{A} \defeq \wt\bigl( \matrH(x) \bigr) \in \Z^{3 \times 4}$, or, let
  $\codeCconv$ be a $(3,4)$-regular type-$2$ convolutional code with
  polynomial parity-check matrix $\matrHconv(y) \in \Ftwoylauser^{3 \times 4}$
  and let $\matr{A} \defeq \wt\bigl( \matrHconv(y) \bigr) \in \Z^{3 \times
    4}$. Then all possible $(3, 4)$-regular size-$(3 \! \times \! 4)$ type-$2$
  weight matrices $\matr{A}$ (up to permutations of rows and columns) are
  given by the following $5$ types of matrices (shown here along with the
  corresponding minimum/free Hamming distance upper bound implied
  by~\eqref{eq:bound:2}):
  \begin{alignat*}{3}
    &
    \begin{bmatrix}
      2 & 2 & 0 & 0 \\
      1 & 1 & 1 & 1 \\
      0 & 0 & 2 & 2
    \end{bmatrix}
    \text{with }
    \{ \dmins, \, \dfree \}
      &\,\leq\,
        &8 + 8 + 8 + 8
     &&= 32, \\   
    &
    \begin{bmatrix}
      2 & 2 & 0 & 0 \\
      1 & 0 & 2 & 1 \\
      0 & 1 & 1 & 2
    \end{bmatrix}
    \text{with }
    \{ \dmins, \, \dfree \}
      &\,\leq\,
         &10 + 10 + 6 + 6
     &&= 32, \\
    &
    \begin{bmatrix}
      2 & 0 & 1 & 1 \\
      1 & 2 & 0 & 1 \\
      0 & 1 & 2 & 1
    \end{bmatrix}
    \text{with }
    \{ \dmins, \, \dfree \}
      &\,\leq\,  
         &7 + 7 + 7 + 9
     &&= 30, \\
    &
    \begin{bmatrix}
      2 & 0 & 1 & 1 \\
      0 & 2 & 1 & 1 \\
      1 & 1 & 1 & 1
    \end{bmatrix}
    \text{with }
    \{ \dmins, \, \dfree \}
      &\,\leq\,  
         &6 + 6 + 8 + 8
     &&= 28, \\
    &
    \begin{bmatrix}
      1 & 1 & 1 & 1 \\
      1 & 1 & 1 & 1 \\
      1 & 1 & 1 & 1 
    \end{bmatrix}
    \text{with }
    \{ \dmins, \, \dfree \}
      &\,\leq\,  
         &6 + 6 + 6 + 6
     &&= 24.
  \end{alignat*}
  As can be seen from this list, the largest upper bound is $\{ \dmins, \,
  \dfree \} \leq 32$ and it can be obtained if the weight matrix $\matr{A}$
  equals (modulo permutations of rows and columns) the first or the second
  matrix in the list.
\end{corollary}

\begin{IEEEproof}
  Omitted.
\end{IEEEproof}

\begin{figure*}
  \begin{center}
    \epsfig{file=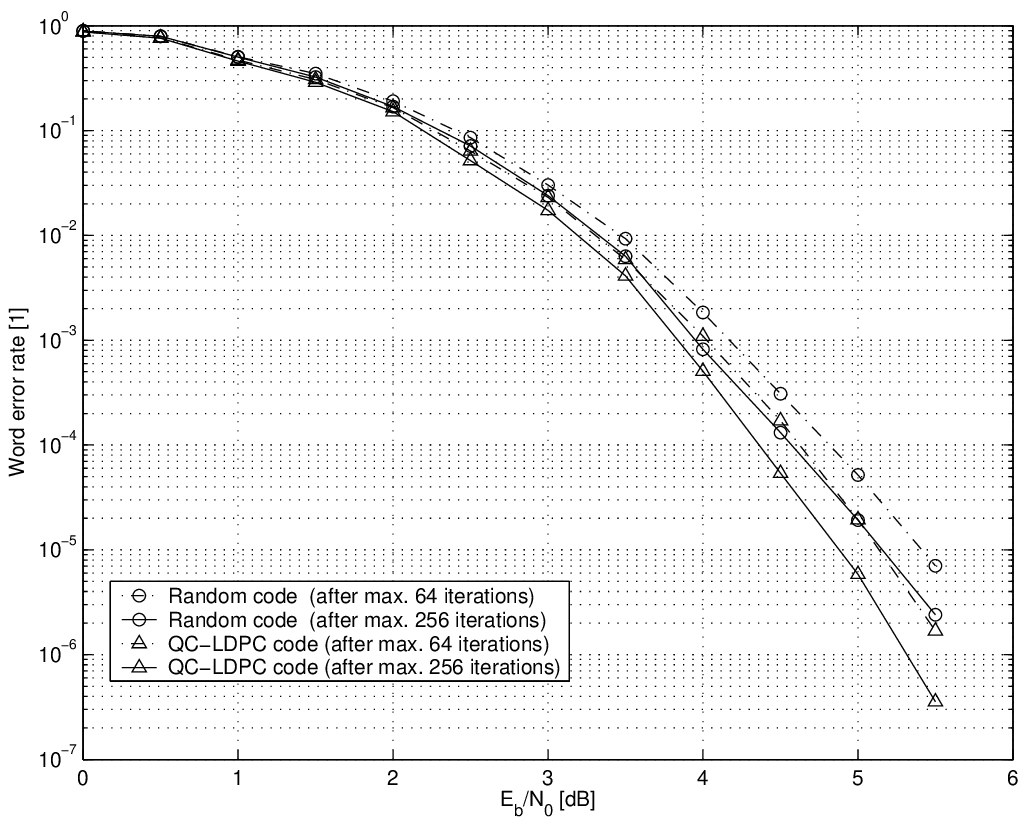, width=0.75\linewidth}
  \end{center}
  \caption{Decoding performance of the $[184,47,32]$ QC LDPC code vs.~a
    randomly generated (four-cycle free) $[184,46]$ LDPC code under
    sum-product algorithm decoding when transmitting over a binary-input AWGN
    channel. (For more details, see Example~\ref{ex:binomial:optimal:2}.)}
  \label{fig:decoding:performance:code:1}
\end{figure*}

\begin{example}
  \label{ex:binomial:optimal:2}

  We see that the type-$2$ $(3,4)$-regular QC code with a $3 \times 4$
  polynomial parity-check matrix and with $r = 46$ presented in
  Example~\ref{ex:binomial:optimal:1} not only achieves the minimum Hamming
  distance upper bound promised by~\eqref{eq:bound:2} but, according to
  Corollary~\ref{cor:type:II:minimum:distance:upper:bound:1}, it achieves the
  best possible minimum Hamming distance upper bound for any type-$2$
  $(3,4)$-regular QC code with a $3 \times 4$ polynomial parity-check matrix.
  We note that this particular code has parameters $[184,47,32]$, girth $8$,
  and diameter $8$, \ie, the same girth and diameter as the Tanner graph of
  the $[124,33,24]$ code in Example~\ref{ex:shortened:tanner:code:1}), which
  is a shortened version of the $[155,64,20]$ code
  in~\cite{Tanner:Sridhara:Fuja:01:1}.
  
  Figure~\ref{fig:decoding:performance:code:1} shows the decoding performance
  (word error rate) of the $[184,47,32]$ QC LDPC code when used for
  transmission over a binary-input additive white Gaussian noise channel.
  Decoding is done using the standard sum-product
  algorithm~\cite{Kschischang:Frey:Loeliger:01} which is terminated if the
  syndrome of the codeword estimate is zero or if a maximal number of $64$
  (respectively~$256$) iterations is reached. It is compared with a randomly
  generated $(3,4)$-regular $[184,46]$ LDPC code where four-cycles in the
  Tanner graph have been eliminated. (When comparing these two codes one has
  to keep in mind that because the randomly generated code has slightly lower
  rate and because the horizontal axis shows $E_{\mathrm{b}}/N_0$, the
  randomly generated code has a slight ``disadvantage'' of $0.093 \
  \mathrm{dB}$.) Note though that the decoding complexity per iteration is the
  same for both codes.
  
  Let us mention on the side that we tried to estimate the minimum (AWGN
  channel) pseudo-weight~\cite{Koetter:Vontobel:03:1,
    Vontobel:Koetter:05:1:subm} of this code. From searching in the
  fundamental cone we get an upper bound of $27.6$ on the minimum
  pseudo-weight. The pseudo-weight spectrum
  gap~\cite{Smarandache:Vontobel:07:1} is therefore estimated to be $32 - 27.6
  = 4.4$, which is on the same order as the pseudo-weight spectrum gap for the
  $(3,5)$-regular $[155,64,20]$ code by
  Tanner~\cite{Tanner:Sridhara:Fuja:01:1}, which is estimated to be $20 - 16.4
  = 3.6$. We also note that for the above-mentioned randomly generated
  $[184,46]$ code we obtained an upper bound on the minimum pseudo-weight of
  $21.0$.
\end{example}

If we want to take into consideration all cases of $(3,4)$-regular
QC/convolutional codes with polynomial parity-check matrices of size $3 \times
4$, we also have to investigate the class of type-$3$ weight matrices of size
$3 \times 4$, as is done in the next corollary.

\begin{corollary}
  \label{cor:type:III:minimum:distance:upper:bound:1}

  Let $\code{C}$ be a $(3,4)$-regular type-$3$ QC code with polynomial
  parity-check matrix $\matrH(x) \in \shortFtwoxmodr^{3 \times 4}$ and let
  $\matr{A} \defeq \wt(\matrH(x)) \in \Z^{3 \times 4}$, or, let $\codeCconv$
  be a $(3,4)$-regular type-$3$ convolutional code with polynomial
  parity-check matrix $\matrHconv(y) \in \Ftwoylauser^{3 \times 4}$ and let
  $\matr{A} \defeq \wt(\matrH(x)) \in \Z^{3 \times 4}$. Then all possible $(3,
  4)$-regular size-$(3 \! \times \! 4)$ type-$3$ weight matrices $\matr{A}$
  (up to permutations of rows and columns) are given by the $5$ types of
  matrices already listed in
  Corollary~\ref{cor:type:II:minimum:distance:upper:bound:1}, together with
  the following $3$ types of matrices (shown here along with the corresponding
  minimum/free Hamming distance upper bound implied by~\eqref{eq:bound:2}):
  \begin{alignat*}{3}     
    &
    \begin{bmatrix}
      3 & 0 & 0 & 1\\
      0 & 2 & 1 & 1\\
      0 & 1 & 2 & 1 
    \end{bmatrix}
    \text{with }
    \{ \dmins, \, \dfree \}
      &\,\leq\,  
         &5 + 9 + 9 + 15
     &&= 38, \\
    &
    \begin{bmatrix}
      3 & 1 & 0 & 0 \\
      0 & 2 & 1 & 1 \\
      0 & 0 & 2 & 2 
    \end{bmatrix}
    \text{with }
    \{ \dmins, \, \dfree \}
      &\,\leq\,
         &4 + 12 + 12 + 12
     &&= 40, \\
    &
    \begin{bmatrix}
      3 & 0 & 0 & 1 \\
      0 & 3 & 0 & 1 \\
      0 & 0 & 3 & 1
    \end{bmatrix} 
    \text{with }
    \{ \dmins, \, \dfree \}
      &\,\leq\,  
         &9 + 9 + 9 + 27
     &&= 54.
  \end{alignat*}
  As it can easily be seen, the largest \emph{upper bound} is $\{ \dmins, \,
  \dfree \} \leq 54$ and it can be obtained if the weight matrix $\matr{A}$
  equals (modulo permutations of rows and columns) the last matrix in the
  above list.
\end{corollary}

\begin{IEEEproof}
  Omitted.
\end{IEEEproof}

\begin{example}
  \label{ex:trinomial:1}

  We can modify the matrix $\matrH$ in
  Example~\ref{ex:shortened:tanner:code:1} to obtain one of the
  configurations in
  Corollary~\ref{cor:type:III:minimum:distance:upper:bound:1}. For example,
  the following matrix ${\matrH}'(x)\in \shortFtwoxmodr^{3\times 4}$
  corresponds to the last configuration listed in
  Corollary~\ref{cor:type:III:minimum:distance:upper:bound:1}:
  \begin{align*}
    \matrH'(x) 
      &\defeq
         \begin{bmatrix}
           x^2{+}x^4{+}x^8  & 0           & 0 & x              \\
           0            & x^9 {+} x^{10}{+}x^{20} & 0 & x^5   \\
           0     & 0 & x^{19} {+} x^7{+}x^{14} & x^{25} 
         \end{bmatrix}.
  \end{align*}
  For $r=31$ we obtain a $[124, 31, 28]$ code, whose rate is $0.25$. (In
  comparison, the monomial $[124, 33, 24]$ QC LDPC code in
  Example~\ref{ex:shortened:tanner:code:1} has rate $0.266$ and the binomial
  $[184,47,32]$ QC LDPC code in Example~\ref{ex:binomial:optimal:1} has rate
  $0.2554$.) For $r = 46$, we obtain a code with parameters $[184,46,34]$. For
  larger $r$ the minimum Hamming distance could increase up to $54$.
\end{example}

Note that the Tanner graph of a polynomial parity-check matrix which has at
least one trinomial entry cannot have girth larger than $6$ (see,
\eg,~\cite{Fossorier:04:1}). We state this observation as part of a more
general analysis of the effect of the weight matrix on the girth.

\begin{theorem}
  \label{th:cycles:patterns:1}

  Let $\cC$ be a QC code described by a polynomial parity-check matrix
  $\matrH(x) \in \shortFtwoxmodr^{\vertsize \times \horsize}$, let $\girth$ be
  the girth of the Tanner graph corresponding to $\matrH(x)$ and let
  $\matr{A}$ be the weight matrix corresponding to $\matrH(x)$. Or, let
  $\codeCconv$ be a convolutional code described by a polynomial parity-check
  matrix $\matrHconv(y) \in \Ftwoylauser^{\vertsize \times \horsize}$, let
  $\girth$ be the girth of the Tanner graph corresponding to $\matrHconv(y)$ and
  let $\matr{A}$ be the weight matrix corresponding to $\matrHconv(y)$.
  \begin{enumerate}

  \item[a)] If $\matr{A}$ has sub-matrix 
    $\begin{bmatrix}
       1 & 1 & 1 \\
       1 & 1 & 1
     \end{bmatrix}$ then $\girth \leq 12$. \\[0.3em]

  \item[b)] If $\matr{A}$ has sub-matrix 
    $\begin{bmatrix}
       1 & 1 \\
       1 & 2
     \end{bmatrix}$ then $\girth \leq 10$. \\[0.5em]

  \item[c)] If $\matr{A}$ has sub-matrix 
    $\begin{bmatrix}
       2 & 2
     \end{bmatrix}$ then $\girth \leq 8$. \\[0.5em]

  \item[d)] If $\matr{A}$ has sub-matrix 
    $\begin{bmatrix}
       3
     \end{bmatrix}$ then $\girth \leq 6$.

  \end{enumerate}
  (By ``$\matr{A}$ having sub-matrix $\matr{B}$'' we mean that $\matr{A}$
  contains a sub-matrix that is equivalent to $\matr{B}$, modulo row
  permutations, column permutations, and transposition.)
\end{theorem}

\begin{IEEEproof}
  See Appendix~\ref{sec:proof:th:cycles:patterns:1}.
\end{IEEEproof}

\mbox{}

The Corollaries~\ref{cor:type:II:minimum:distance:upper:bound:1}
and~\ref{cor:type:III:minimum:distance:upper:bound:1} focused on the case of
$(3,4)$-regular QC/convolutional codes with a $3 \times 4$ polynomial
parity-check matrix. It is clear that similar results can be formulated for
any $(\vertsize ',\horsize')$-regular QC/convolutional code with a
$\vertsize \times \horsize$ polynomial parity-check matrix. However, we will
not elaborate this any further, except for mentioning the following corollary
about $(3,5)$-regular QC/convolutional codes with a $3 \times 5$
polynomial parity-check matrix.

\begin{corollary} 
  An optimal $(3,5)$-regular type-$2$ weight matrix of size $3 \times 5$ must
  (up to row and column permutations) look like
  \begin{align*}
    \matr{A}
      &\defeq
         \begin{bmatrix}
           2 & 2 & 1 & 0 & 0 \\
           0 & 0 & 2 & 2 & 1 \\
           1 & 1 & 0 & 1 & 2
         \end{bmatrix}.
  \end{align*}
  This weight matrix yields the upper bound
  \begin{align*}
    \{ \dmins, \, \dfree \}
      &\leq
         \min\{30, 30, 30, 32, 28 \} 
       = 28.
  \end{align*}
\end{corollary}

\begin{IEEEproof}
  Omitted.
\end{IEEEproof}

\mbox{}

It would be desirable to obtain simply looking bounds for type-$2$ and
type-$3$ codes with $(J,I)$-regular parity-check matrices of size $J \times
I$. Although it is straightforward to obtain such simple bounds by suitably
generalizing the derivation of
Corollary~\ref{cor:type:I:minimum:distance:upper:bound:1}, the resulting
bounds are usually not very useful. We leave it as an open problem to find
such relevant bounds in the style of
Corollary~\ref{cor:type:I:minimum:distance:upper:bound:1} for type-$2$ and
type-$3$ codes.

\section{The Effect of Small Cycles \\
             on the Minimum Hamming Distance \\ 
             and the Free Hamming Distance}
\label{sec:effect:small:cycles:minimum:distance}

If we know that the Tanner graph corresponding to some polynomial parity-check
matrix contains some short cycles then we can strengthen the upper bounds of
Theorem~\ref{th:bound:1} and~\ref{th:bound:2}. In particular,
Theorems~\ref{th:girth:4}, \ref{th:girth:6}, and~\ref{th:girth:gen:1} will
study the influence of $4$-cycles, $6$-cycles, and $2R$-cycles, respectively,
upon the minimum/free Hamming distance upper bounds. These theorems will be
based on results presented in
Lemmas~\ref{lemma:4:cycle:influence:on:permanent:1}
and~\ref{lemma:6:cycle:influence:on:permanent:1} that characterize cycles in
Tanner graphs in terms of some entries of the corresponding polynomial
parity-check matrix, especially in terms of permanents of sub-matrices. In
order to state such conditions, we will use results
from~\cite{Tanner:Sridhara:Fuja:01:1, Fossorier:04:1}. (For other
cycle-characterizing techniques and results, see also~\cite{Wu:You:Zhao:06}
and~\cite{OSullivan:06:1}.)

As we will see, the smaller the girth of the Tanner graph, the smaller the
minimum/free Hamming distance upper bound will be. This observation points in
the same direction as other results do that relate the decoding performance of
LDPC codes (especially under message-passing iterative decoding) to the girth
of their Tanner graph: firstly, there is a lot of empirical evidence that
smaller girth usually hurts the iterative decoding performance; secondly,
there are results concerning the structure of the fundamental polytope that
show that the fundamental polytope of Tanner graphs with smaller girth is
``weaker,'' see, \eg, \cite[Section~8.3]{Vontobel:Koetter:05:1:subm},
\cite{Wainwright:05:1}.

\subsection{Type-I QC/Convolutional Codes with $4$-Cycles}

\begin{lemma}
  \label{lemma:4:cycle:influence:on:permanent:1}

  The Tanner graph of a type-$1$ QC code $\code{C}$ with polynomial
  parity-check matrix $\matrH(x) \in \shortFtwoxmodr^{\vertsize \times
    \horsize}$ has a $4$-cycle if and only if $\matrH(x)$ has a $2 \times 2$
  sub-matrix $\matr{B}(x)$ for which
  \begin{align*}
    \wt\Big( \perm\big( \matr{B}(x) \big) \Big)
      < \perm\Big( \wt\big( \matr{B}(x) \big) \Big)
  \end{align*}
  holds.\footnote{\label{footnote:perm:condition:equivalence:1}Because
    $\wt\big( \perm(\matr{B}(x)) \big) \leq \perm\big( \wt( \matr{B}(x))
    \big)$ for any $\matr{B}(x)$, the condition $\wt\big( \perm(\matr{B}(x))
    \big) < \perm\big( \wt( \matr{B}(x)) \big)$ is equivalent to the condition
    $\wt\big( \perm(\matr{B}(x)) \big) \neq \perm\big( \wt( \matr{B}(x))
    \big)$.}

  An analogous statement can be made for the Tanner graph of a type-$1$
  convolutional code $\codeCconv$ defined by the polynomial parity-check
  matrix $\matrHconv(y) \in \Ftwoylauser^{\vertsize \times \horsize}$.
\end{lemma}

\begin{IEEEproof}
  See Appendix~\ref{sec:proof:lemma:4:cycle:influence:on:permanent:1}.
\end{IEEEproof}

\mbox{}

\begin{corollary}
  \label{cor:on:lemma:4:cycle:influence:on:permanent:1}

  The Tanner graph of a type-$1$ QC code $\code{C}$ with polynomial
  parity-check matrix $\matrH(x) \in \shortFtwoxmodr^{\vertsize \times
    \horsize}$ has a $4$-cycle if and only if $\matrH(x)$ has a $2 \times 2$
  sub-matrix $\matr{B}(x)$ which is monomial and for which $\perm\bigl(
  \matr{B}(x) \bigr) = 0$ (in $\shortFtwoxmodr$) holds.

  An analogous statement can be made for the Tanner graph of a type-$1$
  convolutional code $\codeCconv$ with polynomial parity-check matrix
  $\matrHconv(y) \in \Ftwoylauser^{\vertsize \times \horsize}$.
\end{corollary}

\begin{IEEEproof}
  This follows from Lemma~\ref{lemma:4:cycle:influence:on:permanent:1} and its
  proof.
\end{IEEEproof}

\mbox{}

With this, we are ready to investigate minimum/free Hamming distance upper
bounds for Tanner graphs with $4$-cycles.

\begin{theorem}
  \label{th:girth:4}

  Let $\code{C}$ be a type-$1$ QC code with polynomial parity-check matrix
  $\matrH(x) \in \shortFtwoxmodr^{\vertsize \times \horsize}$, or, let
  $\codeCconv$ be a type-$1$ convolutional code with polynomial parity-check
  matrix $\matrHconv(y) \in \Ftwoylauser^{\vertsize \times \horsize}$. If the
  associated Tanner graph has a $4$-cycle then
  \begin{align}
    \left.
      \begin{array}{c}
        \dmins(\code{C}) \\[0.2cm]
        \dfree(\codeCconv)
      \end{array}
    \right\}
      &\leq
         (\vertsize {+}1)!
         \, - \, 
         2 (\vertsize {-}1)!.
           \label{eq:bound:cycle:1}
   \end{align}
\end{theorem}

\begin{IEEEproof}
  See Appendix~\ref{sec:proof:th:girth:4}.
\end{IEEEproof}

\begin{example}
  Let us consider again the $(2, 4)$-regular length-$4r$ QC code $\code{C}$
  from Example~\ref{ex:bound:1} which is given by the polynomial parity-check
  matrix
  \begin{align*}
    \matrH(x)
      &= \begin{bmatrix}
           x      & x^2   & x^4  & x^8 \\
           x^5    & x^{6} & x^{3} & x^7
         \end{bmatrix}.
  \end{align*}
  It has at least two $4$-cycles since
  \begin{align*}
    \perm
      \left(
        \begin{bmatrix}
          x   & x^2 \\
          x^5 & x^{6}
        \end{bmatrix}
      \right)
      &= 0
    \quad \text{and} \quad
    \perm
      \left(
        \begin{bmatrix}
          x^4   & x^8 \\
          x^{3} & x^7
        \end{bmatrix}
      \right)
       = 0 \; 
  \end{align*}
  (in $\shortFtwoxmodr$). Therefore the bound~\eqref{eq:bound:cycle:1} gives
  \begin{align}
    \dmins(\code{C})
      &\leq
         (\vertsize {+}1)!
         \, - \, 
         2 (\vertsize {-}1)!
       = 3!
         -
         2\cdot 1!
       = 4.
  \end{align}
  We note that for this $\matrH(x)$ this upper bound equals the upper
  bound~\eqref{eq:bound:1} (\confer~Example~\ref{ex:bound:1}).
\end{example}

\subsection{Type-I QC/Convolutional Codes with $6$-Cycles}

\begin{lemma}
  \label{lemma:6:cycle:influence:on:permanent:1}

  The Tanner graph of a type-$1$ QC LDPC code $\code{C}$ with polynomial
  parity-check matrix $\matrH(x) \in \shortFtwoxmodr^{\vertsize \times
    \horsize}$ has a $6$-cycle (or possibly a $4$-cycle) if and only if
  $\matrH(x)$ has a $3 \times 3$ sub-matrix $\matr{B}(x)$ for which
  \begin{align*}
    \wt\Big( \perm\big( \matr{B}(x) \big) \Big)
      &< \perm\Big( \wt\big( \matr{B}(x) \big) \Big)
  \end{align*}
  holds, \ie, if and only if $\matrH(x)$ has a $3 \times 3$ sub-matrix
  $\matr{B}(x)$ for which some terms of its permanent expansion add to
  zero.\footnote{The comment in
    Footnote~\ref{footnote:perm:condition:equivalence:1} applies also here.}

  An analogous statement can be made for the Tanner graph of a type-$1$
  convolutional code $\codeCconv$ defined by the polynomial parity-check
  matrix $\matrHconv(y) \in \Ftwoylauser^{\vertsize \times \horsize}$.
\end{lemma}

\begin{IEEEproof}
  See Appendix~\ref{sec:proof:lemma:6:cycle:influence:on:permanent:1}.
\end{IEEEproof}

\mbox{}

With this, we are ready to investigate the minimum/free Hamming distance upper
bounds for Tanner graphs with $6$-cycles.

\begin{theorem}
  \label{th:girth:6}

  Let $\code{C}$ be a type-$1$ QC code with polynomial parity-check matrix
  $\matrH(x)\in \shortFtwoxmodr^{\vertsize \times \horsize}$, or, let
  $\codeCconv$ be a type-$1$ convolutional code with polynomial parity-check
  matrix $\matrHconv(y)\in \Ftwoylauser^{\vertsize \times \horsize}$. If the
  associated Tanner graph has a $6$-cycle then
  \begin{align*}
    \left.
      \begin{array}{c}
        \dmins(\code{C}) \\[0.2cm]
        \dfree(\codeCconv)
      \end{array}
    \right\}
      \leq
        (\vertsize {+}1)!
        \, - \, 
        2(\vertsize {-}2)!.
   \end{align*}
\end{theorem}

\begin{IEEEproof}
  See Appendix~\ref{sec:proof:th:girth:6}.
\end{IEEEproof}

\mbox{}

\subsection{Type-I QC/Convolutional Codes with $2R$-Cycles}

The previous two subsections have shown that the minimum Hamming distance of a
type-$1$ QC/convolutional code whose Tanner graph has girth $4$ or $6$ can
never attain the maximal value $(\vertsize {+}1)!$ of
Corollary~\ref{cor:type:I:minimum:distance:upper:bound:1}.  These results are
special cases of a more general result that we will discuss next.  Note
however that, compared to the girth-$4$ and the girth-$6$ case, this more
general statement is uni-directional.

\begin{theorem}
  \label{th:girth:gen:1}

  Let $\code{C}$ be a type-$1$ QC code with polynomial parity-check matrix
  $\matrH(x) = [h_{j,i}(x)]_{j,i} \in \shortFtwoxmodr^{\vertsize \times
    \horsize}$. Let $R$, $2 \leq R \leq \min(\vertsize , \horsize )$, be some
  integer, and suppose there is a set $\set{R} \subseteq [\vertsize]$ of size
  $R$, a set $\set{S} \subseteq [\horsize]$ of size $R$, and two distinct
  bijective mappings $\sigma$ and $\tau$ from $\set{R}$ to $\set{S}$ such that
  $\sigma(j) \neq \tau(j)$ for all $j \in \set{R}$ and such that
  \begin{align}
    \prod_{j \in \set{R}}
      h_{j, \sigma(j)}(x)
      &= \prod_{j \in \set{R}}
           h_{j, \tau(j)}(x).
           \label{eq:elementaryproducts:1}
  \end{align} 
  Then
  \begin{align*}
    \dmins
      &\leq
         (\vertsize {+}1)!
         \, - \, 
         2(\vertsize {-}R{+}1)!.
  \end{align*}
  If, in addition, the (bijective) mapping $\sigma^{-1} \circ \tau$ from
  $\set{R}$ to $\set{R}$ is a cyclic permutation of order $R$ and if the
  products on the left-hand and right-hand side of the equation
  in~\eqref{eq:elementaryproducts:1} are nonzero, then the associated Tanner
  graph will have a cycle of length $2 R$.

  An analogous statement can be made for the Tanner graph of a type-$1$
  convolutional code $\codeCconv$ defined by the polynomial parity-check
  matrix $\matrHconv(y) \in \Ftwoylauser^{\vertsize \times \horsize}$.
\end{theorem}

\begin{IEEEproof}
  See Appendix~\ref{sec:proof:th:girth:gen:1}.
\end{IEEEproof} 

\mbox{}

For $4$- and $6$-cycles, the converse of the second part of the above
corollary is true, \ie, $4$- and $6$-cycles are visible in, respectively, $2
\times 2$ and $3 \times 3$ sub-matrices (\confer~Theorems~\ref{th:girth:4}
and~\ref{th:girth:6}). However, for longer cycles the converse of the second
part of the above corollary is \emph{not} always true: $8$-cycles can happen
in $4 \times 4$ sub-matrices, but also in $2 \times 4$ sub-matrices or in $3
\times 4$ sub-matrices. A similar statement holds for longer cycles.

 \subsection{Type-II QC/Convolutional Codes}

With appropriate techniques/computations, similar statements as in the
preceding subsection can also be made about type-$2$ QC/convolutional
codes. We will not say much about this topic except for stating a lemma that
helps in detecting if a polynomial parity-check matrix is $4$-cycle free.

\begin{lemma}
  \label{lemma:four:cycle:free:1}

  A type-$2$ QC code $\code{C}$ is $4$-cycle free if and only if its
  polynomial parity-check matrix $\matrH(x)$ has the following properties.
  \begin{enumerate}

  \item If $r$ is even, then for any $1 \times 1$ sub-matrix like
    \begin{align*}
      \begin{bmatrix}
        x^a + x^b
      \end{bmatrix}
    \end{align*}
    it holds that the permanent of
    \begin{align*}
     \begin{bmatrix}
        x^a & x^b \\
        x^b &  x^a
      \end{bmatrix}
    \end{align*}
    is nonzero (in $\shortFtwoxmodr$). (This condition is equivalent to the
    condition $x^{2a}+x^{2b} \neq 0 \ \text{(in $\Ftwo[x]$)}$, or to the
    condition $\gcd(x^a+x^b, 1+x^r) \neq 1+x^{r/2} \ \text{(in $\Ftwo[x]$)}$.)
 
  \item For any $1 \times 2$ sub-matrix like
    \begin{align*}
      \begin{bmatrix}
        x^a + x^b & x^c + x^d
      \end{bmatrix} \; ,
    \end{align*}
    or any $2 \times 1$ sub-matrix like
    \begin{align*}
      \begin{bmatrix}
        x^a + x^b \\
        x^c + x^d
      \end{bmatrix} \; ,
    \end{align*}
    the product $(x^a + x^b) \cdot (x^c + x^d)$ (in $\shortFtwoxmodr$) has
    weight~$4$, \ie, the maximally possible weight,
    or, equivalently, if all the $2 \times 2$ sub-matrices of the matrix
    \begin{align*}
      \begin{bmatrix}
        x^a & x^b & x^c & x^d \\
        x^b & x^a & x^d & x^c
      \end{bmatrix} \; ,
   \end{align*}
   have nonzero permanent (in $\shortFtwoxmodr$).
 
 \item For any $2 \times 2$ sub-matrix like
    \begin{align*}
      \begin{bmatrix}
        x^a       & x^b + x^c \\
        x^d + x^e & x^f
      \end{bmatrix}
    \end{align*}
    (or row and column permutations thereof), the permanents (in
    $\shortFtwoxmodr$) of the following two $2 \times 2$ sub-matrices
    \begin{align*}
      \begin{bmatrix}
        x^a       & x^b \\
        x^d + x^e & x^f
      \end{bmatrix}
      \text{ \ and \ }
      \begin{bmatrix}
        x^a       & x^c \\
        x^d + x^e & x^f
      \end{bmatrix}
    \end{align*}
    have weight $3$, the maximally possible weight, or, equivalently, if all
    $2 \times 2$ sub-matrices of the matrix
    \begin{align*}
      \begin{bmatrix}
        x^a    &0    & x^b & x^c \\
        0 &x^a& x^c &x^b\\ x^d & x^e & x^f &0\\x^e&x^d&0&x^f
      \end{bmatrix}
    \end{align*}
    have nonzero permanent (in $\shortFtwoxmodr$).

  \item For any $2 \times 2$ sub-matrix with weight matrix
    \begin{align*}
      \begin{bmatrix}
        2 & 2 \\
        1 & 1
      \end{bmatrix},
      \begin{bmatrix}
        2 & 1 \\
        1 & 1
      \end{bmatrix},
      \text{ \ and \ }
      \begin{bmatrix}
        1 & 1 \\
        1 & 1
       \end{bmatrix}
    \end{align*}
    (or row and column permutations thereof), the permanent (in
    $\shortFtwoxmodr$) of this $2 \times 2$ sub-matrix has weight $4$, $3$,
    and $2$, respectively, \ie, the maximally possible weight.
 
  \end{enumerate}

  An analogous statement holds for the Tanner graph of a type-$2$
  convolutional code $\codeCconv$ defined by the polynomial parity-check
  matrix $\matrHconv(y) \in \Ftwoylauser^{\vertsize \times \horsize}$.
\end{lemma}

\begin{IEEEproof}
  It is well known that a $4$-cycle appears in a Tanner graph if and only if
  the corresponding (scalar) parity-check matrix contains the $2 \times 2$
  (scalar) sub-matrix
  \begin{align*}
    \begin{bmatrix}
      1 & 1 \\
      1 & 1
    \end{bmatrix} \; .
  \end{align*}
  The lemma is then proved by studying all possible cases in which a
  polynomial parity-check matrix can lead to a (scalar) parity-check that
  contains this $2 \times 2$ (scalar) sub-matrix. The details are omitted.
\end{IEEEproof}

Note that in the above lemma some of the conditions were expressed in terms of
a double cover of the relevant sub-matrices. In particular, the modified
matrices are obtained by applying the following changes to the entries of the
relevant sub-matrices
\begin{align*}
  x^a + x^b
    &\mapsto
       \begin{bmatrix}
         x^a &x^b \\
         x^b &  x^a
       \end{bmatrix}, \\
  x^f
     &\mapsto 
        \begin{bmatrix}
          x^f &0 \\
          0 &  x^f
        \end{bmatrix}, \\
  0
     &\mapsto 
        \begin{bmatrix}
          \ 0 \ & \ 0 \ \\
          \ 0 \ & \ 0 \ 
        \end{bmatrix}.
\end{align*} 
(Note that similar double covers are also considered in
Appendix~\ref{sec:graph:covers:1}.)

\begin{example}
  Consider, for $r \geq 26$, the type-$2$ polynomial parity-check matrix
  $\matr{H}'(x)$ in~\eqref{eq:binomial:optimal:1}. There, $4$-cycles could
  only be caused by the two sub-matrices
 \begin{align*}
    &
    \begin{bmatrix}
      x^{1} + x^{2} & x^{4}            \\
      x^{5}         & x^{10} + x^{20}
    \end{bmatrix}
    \quad \text{and} \quad
    \begin{bmatrix}
      x^{25} + x^{19} & x^7 + x^{14}
    \end{bmatrix} \; .
 \end{align*}
 Therefore, the conditions for the non-existence of a $4$-cycle are
 \begin{align*}
   0
     &\notin
        \{ 5-1, 5-2 \}
        +
        \{ 4-10, 4-20 \}
     \quad \text{(in $\Zr$)} \; , \\
   0
     &\notin
        \{ 25-19, 19-25 \}
        +
        \{ 7-14 , 14-7 \}
      \quad \text{(in $\Zr$)} \; .
 \end{align*}
 (Here the sum of two sets denotes the set of all possible sums involving one
 summand from the first set and one summand from the second set.) It is clear
 that, with suitable effort, similar analyses could be made for the
 non-existence of longer cycles.
\end{example}

\section{Type-I QC Codes based on \\
               Double Covers of Type-II QC Codes}
\label{sec:type:1:qc:ldpc:codes:based:on:double:cover:1}

So far, we have mostly considered $(\vertsize ,\horsize )$-regular QC codes
that are described by a $\vertsize \times \horsize$ polynomial parity-check
matrix. However, one can construct many interesting $(\vertsize ',
\horsize')$-regular QC LDPC codes with a $\vertsize \times \horsize$
polynomial parity-check matrix where $\vertsize ' \neq \vertsize$ and/or
$\horsize' \neq \horsize$. Given the enormity of the search space, a
worthwhile approach is to start with some small code that has good properties
and to derive longer codes from it. In this section we present such an
approach, along with an analysis of it. Of course, there are many other
possibilities; we leave them open to future studies. (Note that this section
deals only with QC codes, however, similar investigations can also be pursued
for convolutional codes.)

\begin{figure*}
 \begin{center}
   \epsfig{file=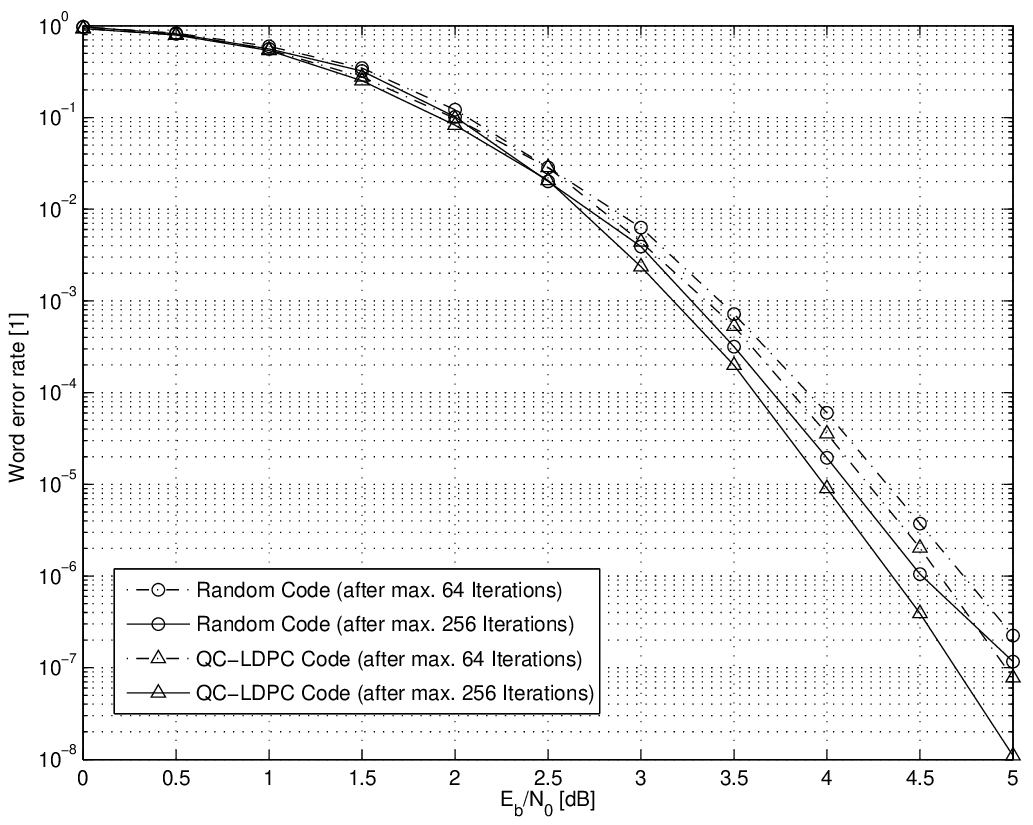, width=0.75\linewidth}
 \end{center}
 \caption{Decoding performance of the $[368,93,32]$ QC LDPC code
   $\code{\tilde{C}}$ and the $[368,93,56]$ QC LDPC code $\code{\hat{C}}$ from
   Example~\ref{monomial:cover:of:binomial} vs.~a randomly generated
   (four-cycle free) $[368, 92]$ LDPC code under sum-product algorithm
   decoding when transmitting over a binary-input AWGN channel. Because the
   performance curve for both QC LDPC codes is nearly the same in the
   simulated signal-to-noise range, we have only shown the performance curve
   of the QC LDPC code $\code{\tilde{C}}$. We observe the onset of an error
   floor of the word error rate at about $4.5$ dB for the randomly generated
   (four-cycle free) LDPC code. A similar observation was made for other
   randomly generated LDPC codes with the same parameters. (Not shown in the
   plot.)}
 \label{fig:decoding:performance:code:2}
\end{figure*}

\begin{example}
  \label{monomial:cover:of:binomial}
 
  Let $\code{C}$ be a QC code described by a $(\vertsize',\horsize')$-regular
  type-$2$ polynomial polynomial parity-check matrix $\matrH(x)$ of size
  $\vertsize \times \horsize$. We would like to derive a type-$1$ polynomial
  parity-check matrix $\matr{\tilde H}(x)$ (of some code $\code{\tilde C}$)
  from $\matrH(x)$. One idea for obtaining such a $\matr{\tilde H}(x)$ is to
  replace all $1 \times 1$ sub-matrices of $\matrH(x)$ by $2 \times 2$
  sub-matrices in the following way:
  \begin{itemize}
  
  \item The sub-matrix $\left[ \begin{smallmatrix} 0 \end{smallmatrix} \right]$
    is replaced by the sub-matrix $\left[ \begin{smallmatrix} 0 & 0 \\ 0 &
        0 \end{smallmatrix} \right]$.
  
  \item A sub-matrix like $\left[ \begin{smallmatrix} x^a \end{smallmatrix}
    \right]$ is replaced by the sub-matrix $\left[ \begin{smallmatrix} x^a & 0
        \\ 0 & x^a \end{smallmatrix} \right]$ (or the sub-matrix $\left[
      \begin{smallmatrix} 0 & x^a \\ x^a & 0 \end{smallmatrix} \right]$).
  
  \item A sub-matrix like $\left[ \begin{smallmatrix} x^a +
        x^b \end{smallmatrix} \right]$ is replaced by the sub-matrix
    $\left[ \begin{smallmatrix} x^a & x^b \\ x^b & x^a \end{smallmatrix}
    \right]$ (or by the sub-matrix $\left[ \begin{smallmatrix} x^b & x^a \\ x^a
        & x^b \end{smallmatrix} \right]$).
  
  \end{itemize}
  Clearly, the resulting matrix $\matr{\tilde H}(x)$ is
  $(\vertsize',\horsize')$-regular and of size $(2\vertsize) \times
  (2\horsize)$, \ie, the same regularity as $\matrH(x)$, but vertically and
  horizontally twice as large as $\matrH(x)$.

  For example, consider the code $\code{C}$ defined by the polynomial
  parity-check matrix $\matrH(x) \in \shortFtwoxmodr^{\vertsize \times
    \horsize}$ in Example~\ref{ex:binomial:optimal:1},\footnote{Note that in
    Example~\ref{ex:binomial:optimal:1} this code was called $\code{C}'$ and
    its polynomial parity-check matrix was called $\matrH'(x)$.} which for
  ease of reference is repeated here
  \begin{align*}
   \matrH(x)
      &\defeq \begin{bmatrix}
           x + x^2 & 0               & x^4             & x^8 \\
           x^5     & x^9             & x^{10} + x^{20}  & 0 \\
           0       & x^{25} + x^{19}  & 0               & x^7 + x^{14}
    \end{bmatrix}.
  \end{align*}
  (Here, $\vertsize = \vertsize ' = 3$ and $\horsize = \horsize' = 4$.)
  Applying the above-mentioned process to $\matrH(x)$ we obtain the following
  type-$1$ $(\tilde \vertsize ', \tilde \horsize')$-regular polynomial
  parity-check matrix $\matr{\tilde H}(x) \in \shortFtwoxmodr^{\tilde
    \vertsize \times \tilde \horsize}$
  \begin{align*}
    \matr{\tilde H}(x)
      &= \begin{bmatrix}
           \begin{matrix}
             x^1 & x^2 \\
             x^2 & x^1
           \end{matrix}
         &
         \begin{matrix} 0 & 0 \\
           0 & 0
         \end{matrix}
         &
         \begin{matrix}
           x^4 & 0 \\
           0          & x^4
         \end{matrix}
         &
         \begin{matrix}
           x^8 & 0 \\
           0          & x^8
         \end{matrix}
         \\
         \begin{matrix}
           x^5 & 0 \\
           0          & x^5
         \end{matrix}
         &
         \begin{matrix}
           x^9 & 0 \\
           0          & x^9
         \end{matrix}
         &
         \begin{matrix}
           x^{10} & x^{20} \\
           x^{20} & x^{10}
         \end{matrix}
         &
         \begin{matrix} 0 & 0 \\
           0 & 0
         \end{matrix}
         \\
         \begin{matrix}
           0 & 0 \\
           0 & 0
         \end{matrix}
         &
         \begin{matrix}
           x^{25} & x^{19} \\
           x^{19} & x^{25}
         \end{matrix}
         &
         \begin{matrix}
           0 & 0 \\
           0 & 0
         \end{matrix}
         &
         \begin{matrix}
           x^7    & x^{14} \\
           x^{14} & x^7
         \end{matrix}
      \end{bmatrix}.
  \end{align*}
  (Here $\tilde \vertsize = 2 \vertsize = 6$, $\tilde \horsize = 2 \horsize =
  8$, $\tilde \vertsize' = \vertsize' = 3$, $\tilde \horsize' = \horsize' =
  4$.)  Clearly, the Tanner graph of $\matr{\tilde H}(x)$ is a double cover of
  the Tanner graph of $\matrH(x)$.\footnote{See
    Appendix~\ref{sec:graph:covers:1} for more details.} Similarly, the
  proto-graph of $\matr{\tilde H}(x)$ is a double cover of the proto-graph of
  $\matrH(x)$.

  For the choice $r = 46$, applying the bounds~\eqref{eq:bound:1}
  and~\eqref{eq:bound:2} to the code $\code{\tilde C}$ described by
  $\matr{\tilde H}(x)$, we obtain, respectively, $\dmins(\code{\tilde C}) \leq
  80$ and $\dmins(\code{\tilde C}) \leq 108$. In addition, from
  $\dmins(\code{C}) = 32$ and
  Lemma~\ref{lemma:double:cover:minimum:distance:1} in
  Appendix~\ref{sec:graph:covers:1} we obtain $32\leq \dmins(\code{\tilde C})
  \leq 2 \dmins(\code{C}) = 2 \cdot 32 = 64$. Moreover, because the matrices
  $\left[ \begin{smallmatrix} 0 & 0 \\ 0 & 0
    \end{smallmatrix} \right]$, $\left[ \begin{smallmatrix} x^a & 0 \\ 0 & x^a
    \end{smallmatrix} \right]$, $\left[ \begin{smallmatrix}  0 & x^a \\ x^a & 0
    \end{smallmatrix} \right]$, $\left[ \begin{smallmatrix} x^a & x^b \\ x^b &
      x^a \end{smallmatrix} \right]$, $\left[ \begin{smallmatrix} x^b & x^a \\
      x^a & x^b \end{smallmatrix} \right]$ commute with each other, and
  because MacKay and Davey's upper bound~\cite{MacKay:Davey:01:1} can be
  reformulated so that it holds also for commuting matrices over polynomials,
  we obtain $\dmins(\code{\tilde C}) \leq 32$. Because the code
  $\code{\tilde{C}}$ has parameters $[368, 93, 32]$, this latter bound
  actually happens to be tight. Therefore, although the above construction can
  produce a new code whose minimum Hamming distance is up to twice as much as
  for the base code, if the base code already reaches the bound given
  by~\eqref{eq:bound:1} then there is no further improvement possible for the
  new code because the above extension of MacKay and Davey's upper bound to
  commuting matrices over polynomials will yield exactly the same upper bound.
 
  However, by changing some entries of polynomial blocks (and thus adding some
  randomness) we can ensure that the nonzero $2\times 2$ polynomial block
  entries do not all commute with each other anymore. Thus, the
  above-mentioned minimum Hamming distance upper bound of $32$ does not apply
  anymore. (In fact, not even the above-mentioned minimum Hamming distance
  upper bound of $64$ applies anymore because the Tanner graph of the modified
  parity-check matrix is not a double cover of the Tanner graph of
  $\matrH(x)$.) Applying such a change to the matrix $\matr{\tilde H}(x)$ (in
  fact, changing only the first block of the matrix $\matr{\tilde H}(x)$) we
  obtain the following $(3,4)$-regular parity-check-matrix
  \begin{align*}
    \matr{\hat{H}}(x)
      &= \begin{bmatrix}
           \begin{matrix}
             x^1 & x^2 \\
             x^2 & x^0
           \end{matrix}
           &
           \begin{matrix} 0 & 0 \\
             0 & 0
           \end{matrix}
           &
           \begin{matrix}
             x^4 & 0 \\
             0          & x^4
           \end{matrix}
           &
           \begin{matrix}
             x^8 & 0 \\
             0          & x^8
           \end{matrix}
           \\
           \begin{matrix}
             x^5 & 0 \\
             0          & x^5
           \end{matrix}
           &
           \begin{matrix}
             x^9 & 0 \\
             0          & x^9
           \end{matrix}
           &
           \begin{matrix}
             x^{10} & x^{20} \\
             x^{20} & x^{10}
           \end{matrix}
           &
           \begin{matrix} 0 & 0 \\
             0 & 0
           \end{matrix}
           \\
           \begin{matrix}
             0 & 0 \\
             0 & 0
           \end{matrix}
           &
           \begin{matrix}
             x^{25} & x^{19} \\
             x^{19} & x^{25}
           \end{matrix}
           &
           \begin{matrix}
             0 & 0 \\
             0 & 0
           \end{matrix}
           &
           \begin{matrix}
             x^7    & x^{14} \\
             x^{14} & x^7
           \end{matrix}
         \end{bmatrix}
  \end{align*}
  for some code $\code{\hat{C}}$. Interestingly enough, for $r = 46$ the code
  $\code{\hat{C}}$ has parameters $[368, 93, 56]$, \ie, the minimum Hamming
  distance is $56$, which is significantly above $32$.\footnote{Note that the
    bounds~\eqref{eq:bound:1} and~\eqref{eq:bound:2} yield, respectively,
    $\dmins(\code{\hat{C}}) \leq 74$ and $\dmins(\code{\tilde C}) \leq 108$.}
  Potentially, the minimum Hamming distance increases even further for
  suitable larger choices of $r$. Note that the rate of $\code{\hat{C}}$ is
  $93/368$, \ie, it is nearly the same as the rate of $\code{C}$, which is
  $47/184$. Of course, the Tanner graph of $\matr{\hat{H}}(x)$ is not a double
  cover of the Tanner graph of $\matr{\tilde H}(x)$, however, its proto-graph
  is a double cover of the proto-graph of $\matr{H}(x)$ and so the Tanner
  graph of $\matr{\hat H}(x)$ is a $2r$-cover of the proto-graph of
  $\matr{H}(x)$.

  The sum-product algorithm decoding performance of codes $\code{\tilde{C}}$
  and $\code{\hat{C}}$ is shown in
  Figure~\ref{fig:decoding:performance:code:2} and compared to a randomly
  generated (four-cycle free) $(3,4)$-regular LDPC code of the same length and
  nearly the same rate. Actually, because the decoding performance of the
  codes $\code{\tilde{C}}$ and $\code{\hat{C}}$ is nearly the same in the
  simulated signal-to-noise range, only the decoding performance of the code
  $\code{\tilde{C}}$ is shown. For higher signal-to-noise ratios and
  correspondingly smaller word error rates we expect that the code
  $\code{\hat{C}}$ will perform better than the code $\code{\tilde{C}}$.
\end{example}

The polynomial parity-check modification methods of this and the previous
sections can now be combined and iterated. For example, one can start with a
type-$1$ polynomial parity-check matrix and form (by rearranging entries) a
type-$2$ or type-$3$ polynomial parity-check matrix. From this, a $2$-cover
type-$1$ matrix (that includes a few twists) can be obtained as discussed
above. Instead of $2$-covers with twists one can also consider $M$-covers with
$M > 2$. For such $M$, the nonzero $M \times M$ sub-matrices that replace the
nonzero $1 \times 1$ sub-matrices can be suitably chosen so that they do not
commute and so that consequently MacKay and Davey's minimum Hamming distance
upper bound does not apply.

This method is just one of many possible ways to construct and analyze a
$(\vertsize ',\horsize')$-regular QC LDPC code with a polynomial parity-check
matrix that has $\vertsize$ rows and $\horsize$ columns with $\vertsize ' \neq
\vertsize$ and/or $\horsize' \neq \horsize$. We leave it for future research to
construct and analyze such codes.

\section{Conclusions}
\label{sec:conclusions:1}

We have presented two minimum/free Hamming distance upper bounds for
QC/convolutional codes, one based on the polynomial parity-check matrix and
one on the weight matrix. Afterwards, we have seen how these upper bounds can
be strengthened based on the knowledge of Tanner graph parameters like
girth. We have also constructed several classes of codes that achieve (or come
close to) these minimum Hamming distance upper bounds. Several extensions of
these results have recently been presented by Butler and
Siegel~\cite{Butler:Siegel:10:1}.

In future work it would be interesting to establish similar bounds for the
minimum pseudo-weight (for different channels) of QC/convolutional LDPC
codes. (Some initial investigations in that direction are presented
in~\cite{Smarandache:Vontobel:09:2}.)

\section*{Acknowledgments}
\label{sec:acknowledgments:1}

We gratefully acknowledge Brian Butler, Oscar Takeshita, and Paul Siegel for
pointing out to us some incompleteness issues with some of the proofs in
earlier versions of this paper. We also gratefully acknowledge discussions
with Irina Bocharova and Florian Hug concerning the minimum Hamming distance
of some of the presented codes and for providing us with the polynomial
parity-check matrix in Example~\ref{example:relatively:high:rate:qc:code:1}.

\appendices

\section{Proof of Theorem~\ref{th:bound:2}}
\label{sec:proof:th:bound:2}

The main part of this appendix is devoted to proving the convolutional code
part of Theorem~\ref{th:bound:2}. The QC code part follows then simply by
combining the convolutional code part of Theorem~\ref{th:bound:2} (applied to
the convolutional code $\codeCconv$ defined by the polynomial parity-check
matrix $\shortmatrHconv(y) \defeq \matrH(x)|_{x = y}$ over $\Ftwoylauser$)
with Tanner's inequality~\eqref{eq:Tanner:dmin:dfree:1}.\footnote{Here and in
  the other appendices, we use the $\shortmatrHconv(y)$ instead of the longer
  $\matrHconv(y)$ for denoting the polynomial parity-check matrix of a
  convolutional code.}

Let us therefore prove the convolutional code part of
Theorem~\ref{th:bound:2}. We use the same notation as in
Lemma~\ref{lemma:codeword:construction:1} and Theorem~\ref{th:bound:1} (and
their proofs). We start by observing that, because the permanent is a certain
sum of certain products, we can use the triangle and product inequality of the
weight function to obtain
\begin{align}
  \wH(\vcconv(y))
    &= \sum_{i \in \set{S}}
         \wt
           \Big(
             \perm\big( \shortmatrHconv_{\set{S} \setminus i}(y) \big)
           \Big) \nonumber \\
    &\leq
       \sum_{i \in \set{S}}
         \perm
           \Big(
             \big(\wt(\shortmatrHconv(y)) \big)_{\set{S} \setminus i}
           \Big) \nonumber \\
    &= \sum_{i \in \set{S}}
         \perm
           \big(
             \matr{A}_{\set{S} \setminus i}
           \big),
             \label{eq:codeword:weight:upper:bound:1}
\end{align}
where in the last step we have used the definition $\matr{A} = \wt\big(
\shortmatrHconv(y) \big)$.

With this, the result in~\eqref{eq:bound:2} is nearly established with the
exception of the case when the codeword construction in
Lemma~\ref{lemma:codeword:construction:1} produces the all-zero vector
$\vcconv(y)$; in Eq.~\eqref{eq:bound:1:convolutional:code} of
Theorem~\ref{th:bound:1} we properly took care of this case by using the
$\minstarnoarg$ operator instead of the $\min$ operator. However, such an
all-zero vector $\vcconv(y)$ can yield a nonzero term $\sum_{i \in \set{S}}
\perm \bigl( \matr{A}_{\set{S} \setminus i} \bigr)$
in~\eqref{eq:codeword:weight:upper:bound:1} and we need to take care of this
degeneracy. Our strategy will be to show that $\dfree(\codeCconv)$ is never
larger than such a nonzero term and therefore, although this nonzero term
appears in the $\minstarnoarg$ operation in~\eqref{eq:bound:2}, it does not
produce a wrong upper bound.

\begin{example}
  Before we continue, let us briefly discuss a polynomial parity-check matrix
  where the above-mentioned degeneracy happens. Let $\code{C}$ be a code with
  polynomial parity-check matrix
  \begin{align*}
    \matrH(x)
      &\defeq
         \begin{bmatrix}
           1 & 1   & 1       & 1       & f(x) \\
           1 & x   & x^2     & x^3     & g(x) \\
           0 & 1+x & 1 + x^2 & 1 + x^3 & h(x)
         \end{bmatrix}
  \end{align*}
  and with weight matrix
  \begin{align*}
    \matr{A}
      &\defeq
         \begin{bmatrix}
           1 & 1 & 1 & 1 & \wt\big( f(x) \big) \\
           1 & 1 & 1 & 1 & \wt\big( g(x) \big) \\
           0 & 2 & 2 & 2 & \wt\big( h(x) \big)
         \end{bmatrix},
  \end{align*}
  where $f(x)$, $g(x)$, and $h(x)$ are some arbitrary polynomials such that
  $h(x) \neq f(x) + g(x)$. The corresponding convolutional code $\codeCconv$
  has parity-check matrix $\shortmatrHconv(y) \defeq \matrH(x)|_{x = y}$.

  Let $\set{S} = \{ 0, 1, 2, 3 \}$. Clearly, the matrix
  $\shortmatrHconv_{\set{S}}(y)$ is rank-deficient because the last row of
  this matrix is the sum of the first two rows. This implies that all $3
  \times 3$ sub-matrices of $\shortmatrHconv_{\set{S}}(y)$ have zero
  determinant, so all $3 \times 3$ sub-matrices of
  $\shortmatrHconv_{\set{S}}(y)$ have zero permanent, and so the codeword
  generating procedure in Lemma~\ref{lemma:codeword:construction:1} yields the
  codeword $\vcconv(y) = (0,0,0,0,0)$ for the above choice of the set
  $\set{S}$. However, the term in~\eqref{eq:codeword:weight:upper:bound:1} is
  \begin{align}
    &
    \sum_{i \in \set{S}}
      \perm
        \big(
          \matr{A}_{\set{S} \setminus i}
        \big) \nonumber \\
      &= \big(
           (2 {\cdot} 1 {\cdot} 1 + 2 {\cdot} 1 {\cdot} 1)
           +
           (2 {\cdot} 1 {\cdot} 1 + 2 {\cdot} 1 {\cdot} 1)
           +
           (2 {\cdot} 1 {\cdot} 1 + 2 {\cdot} 1 {\cdot} 1)
         \big)
         \ + \nonumber \\
      &\quad\ 
         \big(
           (0 {\cdot} 1 {\cdot} 1 + 0 {\cdot} 1 {\cdot} 1)
           +
           (2 {\cdot} 1 {\cdot} 1 + 2 {\cdot} 1 {\cdot} 1)
           +
           (2 {\cdot} 1 {\cdot} 1 + 2 {\cdot} 1 {\cdot} 1)
         \big)
         \ + \nonumber \\
      &\quad\ 
         \big(
           (0 {\cdot} 1 {\cdot} 1 + 0 {\cdot} 1 {\cdot} 1)
           +
           (2 {\cdot} 1 {\cdot} 1 + 2 {\cdot} 1 {\cdot} 1)
           +
           (2 {\cdot} 1 {\cdot} 1 + 2 {\cdot} 1 {\cdot} 1)
         \big)
         \ + \nonumber \\
      &\quad\ 
         \big(
           (0 {\cdot} 1 {\cdot} 1 + 0 {\cdot} 1 {\cdot} 1)
           +
           (2 {\cdot} 1 {\cdot} 1 + 2 {\cdot} 1 {\cdot} 1)
           +
           (2 {\cdot} 1 {\cdot} 1 + 2 {\cdot} 1 {\cdot} 1)
         \big) \nonumber \\
      &= 36.
           \label{eq:example:codeword:weight:upper:bound:1}
  \end{align}
  If we can show that $\dfree(\codeCconv)$ is not larger than $36$ then we are
  sure that the value of $\sum_{i \in \set{S}} \perm \bigl( \matr{A}_{\set{S}
    \setminus i} \bigr)$ does not yield a wrong upper bound. We will do this
  by exhibiting a \emph{nonzero} codeword $\vc'(y)$ with Hamming weight not
  larger than $36$.

  In order to construct such a codeword $\vc'(y)$, let $\matrH'(y)$ be the $2
  \times 5$ sub-matrix of $\shortmatrHconv(y)$ that consists of the first two
  rows of $\shortmatrHconv(y)$, and let $\matr{A}'$ be the $2 \times 5$
  sub-matrix of $\matr{A}$ that consists of the first two rows of
  $\matr{A}$. Because of the above-mentioned rank deficiency of
  $\shortmatrHconv_{\set{S}}(y)$, any vector $\vc'(y)$ in the kernel of
  $\matrH'(y)$ that is zero at the fifth position must be a codeword in
  $\codeCconv$.\footnote{``Fifth position'' refers here to the vector entry
    with index $4$.} Now, applying the codeword generating procedure of
  Lemma~\ref{lemma:codeword:construction:1} for $\matrH'(y)$ and for the set
  $\set{S}' = \{ 0, 1, 2 \}$ we obtain the codeword
  \begin{align*}
    \vc'(y)
      = \begin{pmatrix}
          y+y^2 & 1+y^2 & 1+y & 0 & 0
        \end{pmatrix}.
  \end{align*}
  (Because of the choice of $\set{S}'$, it is clear that the fifth position of
  $\vc'(y)$ is zero. Moreover and most importantly, because the matrix
  $\shortmatrHconv_{\set{S}'}(y)$ has full rank, $\vc'(y)$ is a nonzero
  codeword.\footnote{In the proof of the general case we will also have to
    take into account the case where $\shortmatrHconv_{\set{S}'}(y)$ does
    \emph{not} have full rank.}) This nonzero codeword yields the free Hamming
  distance upper bound
  \begin{align}
    \sum_{i \in \set{S}'}
      \perm
        \big(
          \matr{A}'_{\set{S}' \setminus i}
        \big)
      &= (1 {\cdot} 1 + 1 {\cdot} 1) +
         (1 {\cdot} 1 + 1 {\cdot} 1) +
         (1 {\cdot} 1 + 1 {\cdot} 1) \nonumber \\
      &= 6.
           \label{eq:example:codeword:weight:upper:bound:2}
  \end{align}
  Clearly, $6$ is not larger than $36$, and so the value $\sum_{i \in \set{S}}
  \perm \bigl( \matr{A}_{\set{S} \setminus i} \bigr)$ implied by $\vcconv(y)$
  yields a valid upper bound on the free Hamming distance.

  Alternatively, the fact that the value
  in~\eqref{eq:example:codeword:weight:upper:bound:2} is not larger than the
  value in~\eqref{eq:example:codeword:weight:upper:bound:1} can also be seen
  from the following observation. By multiplying the expression
  in~\eqref{eq:example:codeword:weight:upper:bound:2} by $2$ (the weight of
  the element in the third row and fourth column of $\matr{A}$, \ie, the entry
  of $\matr{A}$ with row index $2$ and column index $3$), we obtain
  \begin{align*}
    (2 {\cdot} 1 {\cdot} 1 + 2 {\cdot} 1 {\cdot} 1) +
    (2 {\cdot} 1 {\cdot} 1 + 2 {\cdot} 1 {\cdot} 1) +
    (2 {\cdot} 1 {\cdot} 1 + 2 {\cdot} 1 {\cdot} 1),
  \end{align*}
  which is a sub-expression
  of~\eqref{eq:example:codeword:weight:upper:bound:1}. Because all terms
  in~\eqref{eq:example:codeword:weight:upper:bound:1} are positive, it is
  clear that the value in~\eqref{eq:example:codeword:weight:upper:bound:2}
  cannot be larger than the value
  in~\eqref{eq:example:codeword:weight:upper:bound:1}.
\end{example}

In order to complete the proof of Theorem~\ref{th:bound:2}, we will generalize
the observations that we have just made in the above example. Let $\set{S}$ be
a subset of $[\horsize]$ with $|\set{S}| = \vertsize + 1$ and let $\vcconv(y)$
be the codeword that is obtained by the codeword generating procedure of
Lemma~\ref{lemma:codeword:construction:1} for the set $\set{S}$. Note that
$\vcconv(y)$ is the all-zero codeword if and only if all $\vertsize \times
\vertsize$ sub-matrices of $\shortmatrHconv_{\set{S}}(y)$ have zero permanent,
if and only if all $\vertsize \times \vertsize$ sub-matrices of
$\shortmatrHconv_{\set{S}}(y)$ have zero determinant, if and only if the
matrix $\shortmatrHconv_{\set{S}}(y)$ is rank-deficient.

We want to show that the value of $\sum_{i \in \set{S}} \perm \bigl(
\matr{A}_{\set{S} \setminus i} \bigr)$, if it is nonzero, is always an upper
bound on $\dfree(\codeCconv)$.

\begin{itemize}

\item Assume that $\shortmatrHconv_{\set{S}}(y)$ has full rank. Then
  $\vcconv(y)$ is a nonzero codeword and so $\sum_{i \in \set{S}} \perm \bigl(
  \matr{A}_{\set{S} \setminus i} \bigr)$ is a free Hamming distance upper
  bound because of the inequalities
  in~\eqref{eq:codeword:weight:upper:bound:1}.

\item Assume that $\shortmatrHconv_{\set{S}}(y)$ has not full rank and that
  $\sum_{i \in \set{S}} \perm \bigl( \matr{A}_{\set{S} \setminus i} \bigr) =
  0$. Then $\perm \bigl( \matr{A}_{\set{S} \setminus i} \bigr) = 0$ for all $i
  \in \set{S}$. It follows that $\perm \bigl( \shortmatrHconv(y)_{\set{S}
    \setminus i} \bigr) = 0$ for all $i \in \set{S}$ and that $\vcconv(y)$ is
  the all-zero codeword. Therefore, although $\vcconv(y)$ is the all-zero
  codeword, this case is properly taken care of by the $\minstarnoarg$
  operator in~\eqref{eq:bound:2}.

\item Finally, assume that $\shortmatrHconv_{\set{S}}(y)$ has not full rank
  and that $\sum_{i \in \set{S}} \perm \bigl( \matr{A}_{\set{S} \setminus i}
  \bigr) > 0$. (Note that this can only happen for $\vertsize \geq 2$.)
  Without loss of generality, we can assume that the rows of
  $\shortmatrHconv(y)$ are ordered such that the last row of
  $\shortmatrHconv_{\set{S}}(y)$ is a linear combination of the first
  $\vertsize - 1$ rows of $\shortmatrHconv_{\set{S}}(y)$. Denote the entries
  of $\shortmatrHconv(y)$ by $h_{j,i}(y)$ and the entries of $\matr{A}$ by
  $a_{j,i}$, and let $\matrH'(y)$ be the $(\vertsize \! - \! 1) \times
  \horsize$ sub-matrix of $\matrH$ consisting of the first $\vertsize - 1$
  rows of $\shortmatrHconv(y)$ and $\matr{A}'$ be the sub-matrix of $\matr{A}$
  that consists of the first $\vertsize - 1$ rows of $\matr{A}$.

  Because of the assumption $\sum_{i \in \set{S}} \perm \bigl(
  \matr{A}_{\set{S} \setminus i} \bigr) > 0$, there must be at least one $i
  \in \set{S}$ such that $\perm \bigl( \matr{A}_{\set{S} \setminus i} \bigr) >
  0$. Using the co-factor expansion of the permanent of $\matr{A}_{\set{S}
    \setminus i}$, this implies that there is at least one $i^{*} \in \set{S}
  \setminus i$ such that
  \begin{align}
    a_{\vertsize-1,i^{*}}
    \cdot
    \perm
      \bigl(
        \matr{A}'_{(\set{S} \setminus i) \setminus i^{*}}
      \bigr)
      &> 0.
           \label{eq:partial:permanent:1}
  \end{align} 
  Fix such an $i^{*}$ and let $\set{S}' \defeq \set{S} \setminus i^{*}$.
  Assume for the moment that $\matrH'_{\set{S}'}(y)$ has full rank. Applying
  the codeword generating procedure in
  Lemma~\ref{lemma:codeword:construction:1} for the polynomial parity-check
  matrix $\matrH'(y)$ and the set $\set{S}'$ we obtain a nonzero vector
  $\vc'(y)$ which is in the kernel of $\matrH'(y)$.  Because of the rank
  deficiency of $\shortmatrHconv_{\set{S}}(y)$ and because $c'_{i'}(y) = 0$
  for $i' \in [\horsize] \setminus \set{S}'$ (and therefore $c'_{i'}(y) = 0$
  for $i' \in [\horsize] \setminus \set{S}$), the vector $\vc'(y)$ must also
  be a codeword in $\codeCconv$. Therefore, because $\vc'(y)$ is a nonzero
  codeword, the free Hamming distance of $\codeCconv$ can be upper bounded as
  follows
  \begin{align*}
    \dfree(\codeCconv)
       \leq \wH(\vc'(y))
      &\leq \sum_{i' \in \set{S}'}
              \perm
                \big(
                  \matr{A}'_{\set{S}' \setminus i'}
                \big) \\
      &\leq a_{\vertsize-1,i^{*}}
            \! \cdot \!\!\!
            \sum_{i' \in \set{S}'}
              \perm
                \big(
                  \matr{A}'_{\set{S}' \setminus i'}
                \big),
  \end{align*}
  where we have used the fact that the inequality
  in~\eqref{eq:partial:permanent:1} implies $a_{\vertsize-1,i^{*}} \geq
  1$. However, because $a_{\vertsize-1,i^{*}} \cdot \sum_{i' \in \set{S}'}
  \perm \bigl( \matr{A}'_{\set{S}' \setminus i'} \bigr)$ is a sub-expression
  of $\sum_{i \in \set{S}} \perm \bigl( \matr{A}_{\set{S} \setminus i} \bigr)$
  we obtain
  \begin{align*}
    \dfree(\codeCconv)
      &\leq \sum_{i \in \set{S}}
              \perm
                \big(
                  \matr{A}_{\set{S} \setminus i}
                \big),
  \end{align*}
  where we have used the fact that $\sum_{i \in \set{S}} \perm \bigl(
  \matr{A}_{\set{S} \setminus i} \bigr)$ contains only non-negative terms.

  It remains the case where $\matrH'_{\set{S}'}(y)$ has not full rank. We can
  solve this case with a similar procedure as above. Note that the above
  choice of $i^{*}$ ensures that there will be a suitable $i'^{*} \in
  \set{S}'$.
\end{itemize}

\section{Proof of Corollary~\ref{cor:type:I:minimum:distance:upper:bound:1}}
\label{sec:proof:cor:type:I:minimum:distance:upper:bound:1}

The main part of this appendix is devoted to proving the convolutional code
part of Corollary~\ref{cor:type:I:minimum:distance:upper:bound:1}. The QC code
part follows then simply by combining the convolutional code part of
Corollary~\ref{cor:type:I:minimum:distance:upper:bound:1} (applied to the
convolutional code $\codeCconv$ defined by the polynomial parity-check matrix
$\shortmatrHconv(y) \defeq \matrH(x)|_{x = y}$ over $\Ftwoylauser$) with
Tanner's inequality~\eqref{eq:Tanner:dmin:dfree:1}.

Let us therefore prove the convolutional code part of
Corollary~\ref{cor:type:I:minimum:distance:upper:bound:1}. We have to consider
two cases. First, assume that the weight matrix $\matr{A}$ is such that there
is at least one set $\set{S}' \subseteq [\horsize]$ with $|\set{S}'| =
\vertsize +1$ such that $\sum_{i \in \set{S}'} \perm \left( \matr{A}_{\set{S}'
    \setminus i} \right) > 0$. Using the fact that the weight matrix
$\matr{A}$ of a type-$1$ convolutional code contains only zeros and ones, we
can conclude that for such an $\set{S}'$ we have $\perm \left(
  \matr{A}_{\set{S}' \setminus i} \right) \leq \vertsize !$ for all $i \in
\set{S}'$, which implies that
\begin{align*}
  \dfree(\codeCconv)
    &\leq
       \minstar{\set{S} \subseteq [\horsize] 
                \atop |\set{S}| = \vertsize +1} \ 
         \sum_{i \in \set{S}} \ 
           \perm
             \left(
               \matr{A}_{\set{S} \setminus i}
             \right) \\
    &\leq
       \sum_{i \in \set{S}'} \ 
         \perm
           \left(
             \matr{A}_{\set{S}' \setminus i}
           \right) \\
    &\leq
       \sum_{i \in \set{S}'} \ 
           \vertsize  !
     \ = \ (\vertsize +1) \cdot \vertsize  !
     \ = \ (\vertsize +1) ! .
\end{align*}
This is the upper bound that we set out to prove.

Secondly, assume that the weight matrix $\matr{A}$ is such that for all sets
$\set{S} \subseteq [\horsize]$ with $|\set{S}| = \vertsize +1$ it holds that
$\sum_{i \in \set{S}} \perm \left( \matr{A}_{\set{S} \setminus i} \right) =
0$. Notably, this implies that $\perm \left( \matr{A}_{\set{S} \setminus i}
\right) = 0$ for all sets $\set{S}$ and all $i \in \set{S}$. (Parity-check
matrices with such a weight matrix $\matr{A}$ are rather degenerate and in
general uninteresting. However, we need to properly take care of this case too
in order to verify that the corollary statement holds for all possible
type-$1$ convolutional codes.) From the above condition it follows that
$\shortmatrHconv(y)$ must be such that for all sets $\set{S} \subseteq
[\horsize]$ with $|\set{S}| = \vertsize +1$ and for all $i \in \set{S}$ it
holds that $\perm \left( \shortmatrHconv_{\set{S} \setminus i}(y) \right) =
0$, \ie, $\det \left( \shortmatrHconv_{\set{S} \setminus i}(y) \right) =
0$. This latter statement, however, is equivalent to the statement that
$\shortmatrHconv(y)$ does not have full row rank. The code $\codeCconv$ can
therefore also be defined by a suitably chosen $(\vertsize {-}1) \times
\horsize$ sub-matrix of $\shortmatrHconv(y)$. If $\vertsize > 1$ then applying
this corollary recursively to this $(\vertsize {-}1) \times \horsize$
sub-matrix we obtain $\dfree(\codeCconv) \leq ((\vertsize {-}1){+}1)!  =
\vertsize !$, which implies $\dfree(\codeCconv) \leq (\vertsize
{+}1)!$. Otherwise (\ie, when $\vertsize = 1$), it clearly holds that
$\dfree(\codeCconv) \leq (\vertsize {+}1)! = 2!  = 2$.

\section{Proof of Theorem~\ref{th:cycles:patterns:1}}
\label{sec:proof:th:cycles:patterns:1}

We prove only the QC code part of the theorem. The convolutional code part
follows by a similar argument.

Assume that $\cC$ has polynomial parity-check matrix $\matrH(x)$. We establish
upper bounds on the girth of the Tanner graph of $\matrH(x)$ by exhibiting the
existence of certain cycles in that graph. These cycles are found using
techniques from~\cite{Tanner:Sridhara:Fuja:01:1, Fossorier:04:1}.
\begin{enumerate}

\item[a)] Let
  \begin{align*}
    \begin{bmatrix}
      x^a & x^b & x^c \\
      x^d & x^e & x^f
    \end{bmatrix}
  \end{align*}
  be any sub-matrix of $\matrH(x)$ having the first weight configuration. The
  path
  \begin{align*}
    x^a
      &\rightarrow
         x^d
       \rightarrow
         x^f
       \rightarrow
         x^c
       \rightarrow
         x^b
       \rightarrow
         x^e
       \rightarrow
         x^d
       \rightarrow
         x^a \\
      &\rightarrow
         x^ c
       \rightarrow
         x^f
       \rightarrow
         x^e
       \rightarrow
         x^b
       \rightarrow
         x^a,
  \end{align*}
  shows that the Tanner graph of $\matrH(x)$ has at least one $12$-cycle
  since
  \begin{align*}
    &
    (a{-}d) + 
    (f{-}c) + 
    (b{-}e) +
    (d{-}a) \\
    &
    + 
    (c{-}f) + 
    (e{-}b)
       = 0
  \end{align*}
  in $\Z$ (and therefore also in $\Zr$).
 
\item[b)] Let
  \begin{align*}
    \begin{bmatrix}
      x^a & x^c \\
      x^b & x^d + x^e
    \end{bmatrix}
  \end{align*}
  be a sub-matrix matrix of $\matrH(x)$ having the second weight
  configuration. The path
  \begin{align*}
    x^a
      &\rightarrow
         x^b
       \rightarrow 
         x^d
       \rightarrow
         x^e
       \rightarrow 
         x^b
       \rightarrow 
         x^a \\
      &\rightarrow
         x^c 
       \rightarrow 
         x^d 
       \rightarrow 
         x^e
       \rightarrow
         x^c
       \rightarrow 
         x^a
  \end{align*}
  shows that the Tanner graph of $\matrH(x)$ has at least one $10$-cycle
  since
  \begin{align*}
    (a{-}b) +
    (d{-}e) + 
    (b{-}a) + 
    (c{-}d) + 
    (e{-}c)
      &= 0
  \end{align*}
  in $\Z$ (and therefore also in $\Zr$).

\item[c)] Let
  \begin{align*}
    \begin{bmatrix}
      x^a + x^b & x^c + x^d
    \end{bmatrix}
  \end{align*}
  be a sub-matrix matrix of $\matrH$ having the third weight configuration.
  The path
  \begin{align*}
    x^a
      &\rightarrow 
         x^b 
       \rightarrow 
         x^c 
       \rightarrow
         x^d
       \rightarrow 
         x^b
       \rightarrow 
         x^a
       \rightarrow
         x^d 
       \rightarrow 
         x^c 
       \rightarrow
         x^a
  \end{align*}
  shows that the Tanner graph of $\matrH(x)$ has at least one $8$-cycle
  since
  \begin{align*}
    (a{-}b) +
    (c{-}d) +
    (b{-}a) +
    (d{-}c)
      &= 0
  \end{align*}
  in $\Z$ (and therefore also in $\Zr$).

\item[d)] Let
  \begin{align*}
    \begin{bmatrix}
      x^a + x^b + x^c
    \end{bmatrix}
  \end{align*}
  be a sub-matrix matrix of $\matrH$ having the stated weight configuration.
  The path
  \begin{align*}
    x^a
      &\rightarrow 
         x^b 
       \rightarrow 
         x^c 
       \rightarrow
         x^a
       \rightarrow 
         x^b
       \rightarrow 
         x^c
       \rightarrow
         x^a
  \end{align*}
  shows that the Tanner graph of $\matrH(x)$ has at least one $6$-cycle
  since
  \begin{align*}
    (a{-}b) + 
    (c{-}a) + 
    (b{-}c)
      &= 0
  \end{align*}
  in $\Z$ (and therefore also in $\Zr$).

\end{enumerate}

\section{Proof of 
               Lemma~\ref{lemma:4:cycle:influence:on:permanent:1}}
\label{sec:proof:lemma:4:cycle:influence:on:permanent:1}

We prove only the QC code part of the lemma. The convolutional code part
follows by a similar argument.

Note that a $2 \times 2$ sub-matrix $\matr{B}(x)$ must, up to row and column
permutations, look
\begin{alignat*}{2}
  \text{either like }
  &
  \begin{bmatrix}
    x^a & x^b \\
    x^c & x^d
  \end{bmatrix}
  \text{ or like }
  &
  \begin{bmatrix}
    x^a & x^b \\
    0   & x^d
  \end{bmatrix}
  \text{ or like }
  &
  \begin{bmatrix}
    x^a & 0   \\
    0   & x^d
  \end{bmatrix} \\
  \text{ or like }
  &
  \begin{bmatrix}
    x^a & 0 \\
    x^c & 0
  \end{bmatrix}
  \text{ or like }
  &
  \begin{bmatrix}
    x^a & x^b \\
    0   & 0
  \end{bmatrix}
  \text{ or like }
  &
  \begin{bmatrix}
    x^a & 0   \\
    0   & 0
  \end{bmatrix},
\end{alignat*}
for some $a, b, c, d \in \Zr$.

In the first case,
\begin{align*}
  \wt\Big( \perm\big( \matr{B}(x) \big) \Big)
    &< \perm\Big( \wt\big( \matr{B}(x) \big) \Big)
\end{align*}
holds if and only if
\begin{align*}
  \wt\left( x^{a+d} + x^{b+c} \right)
    &< 2,
\end{align*}
if and only if
\begin{align*}
  x^{a+d} + x^{b+c}
    &= 0 \quad \text{(in $\shortFtwoxmodr$)},
\end{align*}
if and only if
\begin{align*}
  a + d
    &= b + c \quad \text{(in $\Zr$)},
\end{align*}
if and only if
\begin{align*}
  a - c + d - b
     = 0 \quad \text{(in $\Zr$)},
\end{align*} 
which is equivalent to the existence of a $4$-cycle in the Tanner graph, see
the conditions in~\cite{Tanner:Sridhara:Fuja:01:1, Fossorier:04:1}.

In the second and third case, $\wt\bigl( \perm(\matr{B}(x)) \bigr) <
\perm\bigl( \wt(\matr{B}(x)) \bigr)$ holds if and only if $\wt\bigl( x^{a+d}
\bigr) < 1$, \ie, if and only if $1 < 1$. However, this is never the case.
This agrees with the observation that such $2 \times 2$ sub-matrices cannot
induce a four-cycle in the Tanner graph~\cite{Tanner:Sridhara:Fuja:01:1,
  Fossorier:04:1}.

In the fourth, fifth, and sixth case, $\wt\bigl( \perm(\matr{B}(x)) \bigr) <
\perm\bigl( \wt(\matr{B}(x)) \bigr)$ holds if and only if $\wt(0) < 0$, \ie,
if and only if $0 < 0$. However, this is never the case. This agrees with the
observation that such $2 \times 2$ sub-matrices cannot induce a four-cycle in
the Tanner graph~\cite{Tanner:Sridhara:Fuja:01:1, Fossorier:04:1}.

The proof is concluded by noting that a $4$-cycle can appear only in a $2
\times 2$ sub-matrix of a type-$1$ polynomial parity-check matrix.

\section{Proof of Theorem~\ref{th:girth:4}}
\label{sec:proof:th:girth:4}

The main part of this appendix is devoted to proving the convolutional code part
of Theorem~\ref{th:girth:4}. The QC code part will be considered at the end of
this appendix.

The proof of this theorem is based on upper bounding the free Hamming distance
upper bound in~\eqref{eq:bound:1:convolutional:code}, thereby taking advantage
of the fact that $\shortmatrHconv(y) \defeq \matrHconv(y)$ is assumed to be of
type $1$ and to have a $4$-cycle. For ease of reference,
Eq.~\eqref{eq:bound:1:convolutional:code} is repeated here, \ie,
\begin{align}
  \dfree(\codeCconv)
    &\leq
       \minstar{\set{S} \subseteq [\horsize] \atop |\set{S}| = \vertsize +1} \ 
         \sum_{i \in \set{S}}
           \wt
             \Big(
               \perm\big( \shortmatrHconv_{\set{S} \setminus i}(y) \big)
             \Big).
               \label{eq:bound:1:convolutional:code:repetition:1}
\end{align}
Without loss of generality, we can assume that the rows and columns of
$\shortmatrHconv(y)$ are labeled such that the present $4$-cycle implies that
the sub-block
\begin{align}
  \begin{bmatrix}
    h_{00}(y) & h_{01}(y) \\
    h_{10}(y) & h_{11}(y)
  \end{bmatrix}
\end{align}
has permanent $0$ (in $\Ftwoylauser$), \ie, that
\begin{align}
  h_{00}(y)
  h_{11}(y)
  +
  h_{01}(y)
  h_{10}(y)
    &= 0
       \quad \text{(in $\Ftwoylauser$)}.
    \label{eq:proof:th:girth:4:assumption:y:1}
\end{align}

We define the following sets.
\begin{itemize}

\item Let $S_{\nsupseteq \{ 0, 1 \}}$ be the set of all sets $\set{S}
  \subseteq [\horsize]$ with $|\set{S}| = \vertsize {+}1$ that are \emph{not}
  supersets of $\{ 0, 1 \}$.

\item Let $S_{\supseteq \{ 0, 1 \}}$ be the set of all sets $\set{S} \subseteq
  [\horsize]$ with $|\set{S}| = \vertsize {+}1$ that are supersets of $\{ 0, 1
  \}$.

\end{itemize}
Then~\eqref{eq:bound:1:convolutional:code:repetition:1} can be rewritten to
read
\begin{align}
  \dfree(\codeCconv)
    &\leq
       \min
         \bigg(
           \minstar{\set{S} \in S_{\nsupseteq \{ 0, 1 \}}}
             \sum_{i \in \set{S}}
               \wt
                 \Big(
                   \perm\big( \shortmatrHconv_{\set{S} \setminus i}(y) \big)
                 \Big), 
                   \nonumber \\
    &\quad\quad\quad\quad
           \minstar{\set{S} \in S_{\supseteq \{ 0, 1 \}}}
             \sum_{i \in \set{S}}
               \wt
                 \Big(
                   \perm\big( \shortmatrHconv_{\set{S} \setminus i}(y) \big)
                 \Big)
         \bigg).
               \label{eq:bound:1:convolutional:code:repetition:1:mod:2}
\end{align}
The first argument of the min-operator
in~\eqref{eq:bound:1:convolutional:code:repetition:1:mod:2} can be addressed
with a reasoning that is akin to the reasoning in the proofs of
Theorem~\ref{th:bound:2} (\confer~Appendix~\ref{sec:proof:th:bound:2}) and
Corollary~\ref{cor:type:I:minimum:distance:upper:bound:1}
(\confer~Appendix~\ref{sec:proof:cor:type:I:minimum:distance:upper:bound:1}).
This yields
\begin{align}
  \minstar{\set{S} \in S_{\nsupseteq \{ 0, 1 \}}} \ 
    \sum_{i \in \set{S}}
      \wt
        \Big(
          \perm\big( \shortmatrHconv_{\set{S} \setminus i}(y) \big)
        \Big)
    &\leq
       (\vertsize +1)!.
         \label{eq:bound:1:convolutional:code:repetition:1:mod:3}
\end{align}

Therefore, let us focus on the second argument of the min-operator
in~\eqref{eq:bound:1:convolutional:code:repetition:1:mod:2}, \ie,
\begin{align}
  \minstar{\set{S} \in S_{\supseteq \{ 0, 1 \}}} \ 
    \sum_{i \in \set{S}}
      \wt
        \Big(
          \perm\big( \shortmatrHconv_{\set{S} \setminus i}(y) \big)
        \Big).
         \label{eq:bound:1:convolutional:code:repetition:1:mod:4}
\end{align}
We consider two sub-cases. First, assume that the polynomial parity-check
matrix $\shortmatrHconv(y)$ is such that there is at least one set $\set{S}'
\in S_{\supseteq \{ 0, 1 \}}$ such that $\sum_{i \in \set{S}'} \wt \Big(
\perm\big( \shortmatrHconv_{\set{S}' \setminus i}(y) \big) \Big) > 0$. Any
upper bound on this sum will be a valid upper bound on the expression
in~\eqref{eq:bound:1:convolutional:code:repetition:1:mod:4}. For $i \in \{ 0,
1 \}$ we find that
\begin{align}
  \wt
    \Big(
      \perm\big( \shortmatrHconv_{\set{S}' \setminus i}(y) \big)
    \Big)
    &\leq \vertsize !,
       \label{eq:proof:th:girth:4:case:2:1}
\end{align}
where we used the fact that $\shortmatrHconv(y)$ is a type-$1$ polynomial
parity-check matrix.

For $i \in \set{S}' \setminus \{ 0, 1 \}$, however, we want to use a more
refined analysis. We define the following sets.
\begin{itemize}

\item We define $\set{P}'$ to be the set of all permutation mappings from
  $[\vertsize]$ to $\set{S}' \setminus \{ i \}$.

\item We define $\set{P}'' \subseteq \set{P}'$ to be the set of all
  permutation mappings from $[\vertsize]$ to $\set{S}' \setminus \{ i \}$ that
  map $(0, 1)$ to $(0, 1)$ or map $(0, 1)$ to $(1, 0)$.

\item We define $\set{P}''_{\{ 0, 1 \}}$ to be the set of all permutation
  mappings from $[\vertsize] \setminus \{ 0, 1 \}$ to $\set{S}' \setminus \{
  0, 1, i \}$.

\end{itemize}
With these definitions, we obtain
\begin{align*}
  &
  \perm\big( \shortmatrHconv_{\set{S}' \setminus i}(y) \big) \\
    &= \sum_{\sigma \in \set{P}'}
         \prod_{j \in [\vertsize]}
           h_{j, \sigma (j)}(y) \\
    &= \sum_{\sigma \in \set{P}''}
         \prod_{j \in [\vertsize]}
           \!
           h_{j, \sigma (j)}(y)
       +
       \sum_{\sigma \in \set{P}' \setminus \set{P}''}
         \prod_{j \in [\vertsize]}
           \!\!
           h_{j, \sigma (j)}(y) \\
    &= \big(
         h_{00}(y)
         h_{11}(y)
         {+}
         h_{01}(y)
         h_{10}(y)
       \big)
       \cdot\ 
       \!\!\!\!\!\!\!
       \sum_{\sigma \in \set{P}''_{\{ 0, 1 \}}}
         \prod_{j \in [\vertsize] \setminus \{ 0, 1 \}}
           \!\!\!\!\!\!\!
           h_{j, \sigma (j)}(y) \\
    &\quad\
       +
       \sum_{\sigma \in \set{P}' \setminus \set{P}''}
         \prod_{j \in [\vertsize]}
           h_{j, \sigma (j)}(y) \\
    &= \sum_{\sigma \in \set{P}' \setminus \set{P}''}
         \prod_{j \in [\vertsize]}
           h_{j, \sigma (j)}(y)
       \quad \text{(in $\Ftwoylauser$)},
\end{align*}
where in the last equality we have taken advantage
of~\eqref{eq:proof:th:girth:4:assumption:y:1}. Clearly, $|\set{P}' \setminus
\set{P}''| = |\set{P}'| - |\set{P}''| = \vertsize ! - 2(\vertsize-2)!$, so
that we can upper bound the weight of $\perm\bigl( \shortmatrHconv_{\set{S}' \setminus
  i}(y) \bigr)$ as follows
\begin{align}
  \wt
    \Big(
      \perm\big( \shortmatrHconv_{\set{S}' \setminus i}(y) \big)
    \Big)
    &\leq 
       \vertsize !
       \, - \, 
       2 (\vertsize {-}2)!,
       \label{eq:proof:th:girth:4:case:2:2}
\end{align}
where we have again used the fact that $\shortmatrHconv(y)$ is a type-$1$
polynomial parity-check matrix. Combining~\eqref{eq:proof:th:girth:4:case:2:1}
and~\eqref{eq:proof:th:girth:4:case:2:2}, we obtain
\begin{align}
  \sum_{i \in \set{S}'}
    \wt
      &
      \Big(
        \perm\big( \shortmatrHconv_{\set{S}' \setminus i}(y) \big)
      \Big) \nonumber \\
    &= \sum_{i \in \{ 0, 1 \}}
         \wt
           \Big(
             \perm\big( \shortmatrHconv_{\set{S}' \setminus i}(y) \big)
           \Big) \nonumber \\
    &\quad\ 
       +
       \sum_{i \in \set{S}' \setminus \{ 0, 1 \}}
         \wt
           \Big(
             \perm\big( \shortmatrHconv_{\set{S}' \setminus i}(y) \big)
           \Big) \nonumber \\
    &\leq
       2 \vertsize !
       +
       (\vertsize {-}1) (\vertsize ! - 2 (\vertsize {-}2)!) \nonumber \\
    &\leq 
       (\vertsize {+}1)!
       \, - \, 
       2 (\vertsize {-}1)!.
       \label{eq:proof:th:girth:4:case:2:3}
\end{align}

It remains to address the second sub-case, namely where we assume that the
polynomial parity-check matrix $\shortmatrHconv(y)$ is such that for all sets
$\set{S} \in S_{\supseteq \{ 0, 1 \}}$ it holds that $\sum_{i \in \set{S}} \wt
\Big( \perm\big( \shortmatrHconv_{\set{S} \setminus i}(y) \big) \Big) =
0$. This, however, is equivalent to the assumption that for all sets $\set{S}
\in S_{\supseteq \{ 0, 1 \}}$ and all $i \in \set{S}$ it holds that
$\perm\big( \shortmatrHconv_{\set{S} \setminus i}(y) \big) = \det\big(
\shortmatrHconv_{\set{S} \setminus i}(y) \big) = 0$, which in turn is
equivalent to the assumption that for all $\set{S} \in S_{\supseteq \{ 0, 1
  \}}$ the sub-matrix $\shortmatrHconv_{\set{S}}(y)$ does not have full row
rank.

Pick any $\set{S}' \in S_{\supseteq \{ 0, 1 \}}$ and let code $\codeCconv'$ be
the code defined by $\shortmatrHconv_{\set{S}'}(y)$. Without loss of
generality, we can assume that the rows of $\shortmatrHconv(y)$ are ordered
such that the last row of $\shortmatrHconv_{\set{S}'}(y)$ is a linear
combination of the first $\vertsize - 1$ rows of
$\shortmatrHconv_{\set{S}'}(y)$. Let $\codeCconv''$ be the code that is
defined by the $(\vertsize \!  - \! 1) \times (\vertsize \! + \! 1)$
sub-matrix $\shortmatrHconv_{[\vertsize-1],\set{S}'}(y)$ of
$\shortmatrHconv_{\set{S}'}(y)$. Clearly,
\begin{align*}
  \dfree(\codeCconv)
    &\leq
       \dfree(\codeCconv')
     = \dfree(\codeCconv'') \\
    &\leq
       \big(
         (\vertsize {-}1)+1
       \big)!
     = \vertsize !,
\end{align*}
where the first step follows from the fact that any nonzero codeword in
$\codeCconv'$ induces a nonzero codeword in $\codeCconv$, the second step
follows from the equivalence of $\codeCconv'$ and $\codeCconv''$, and the
third step follows from
Corollary~\ref{cor:type:I:minimum:distance:upper:bound:1}. Without loss of
generality we can assume that $\vertsize \geq 2$ (otherwise a type-$1$
polynomial parity-check matrix $\shortmatrHconv(y)$ cannot have a four-cycle),
and so we can upper bound the previous result as follows
\begin{align}
  \dfree(\codeCconv)
    &\leq
       (\vertsize {+}1)!
       \, - \, 
       2 (\vertsize {-}1)!.
       \label{eq:proof:th:girth:4:case:2:4}
\end{align}

Finally, combining~\eqref{eq:bound:1:convolutional:code:repetition:1:mod:3}
and~\eqref{eq:proof:th:girth:4:case:2:3},
or~\eqref{eq:bound:1:convolutional:code:repetition:1:mod:3}
and~\eqref{eq:proof:th:girth:4:case:2:4}, we conclude that the convolutional
code part of Theorem~\ref{th:girth:4} is indeed correct, independently of
which sub-case applies.

The QC code part of Theorem~\ref{th:girth:4} can now be obtained as
follows. Without loss of generality, we can assume that the rows and columns
of $\matrH(x)$ are labeled such that the present $4$-cycle implies that the
sub-block
\begin{align*}
  \begin{bmatrix}
    h_{00}(x) & h_{01}(x) \\
    h_{10}(x) & h_{11}(x)
  \end{bmatrix}
\end{align*}
has permanent $0$ (in $\shortFtwoxmodr$), \ie, that
\begin{align*}
  h_{00}(x)
  h_{11}(x)
  +
  h_{01}(x)
  h_{10}(x)
    &= 0
       \quad \text{(in $\Ftwoxmodr$)}.
\end{align*}
Note that this does \emph{not}
imply~\eqref{eq:proof:th:girth:4:assumption:y:1} for $\shortmatrHconv(y)
\defeq \matrH(x)|_{x = y}$. However, because multiplying a row of $\matrH(x)$
by an invertible element of $\Ftwoxmodr$ produces a parity-check matrix for
the same code, and because multiplying a column of $\matrH(x)$ by a monomial
produces a parity-check matrix of an equivalent code,\footnote{Two binary
  codes are called equivalent if the two codeword sets are equal (up to
  coordinate permutations). Clearly, equivalent codes have the same minimum
  Hamming distance.} we can, without loss of generality, assume that
$\matrH(x)$ is such that
\begin{align*}
  \begin{bmatrix}
    h_{00}(x) & h_{01}(x) \\
    h_{10}(x) & h_{11}(x)
  \end{bmatrix}
    &= \begin{bmatrix}
         1 & 1 \\
         1 & 1
  \end{bmatrix}.
\end{align*}
For such a reformulated $\matrH(x)$, the polynomial parity-check matrix
$\shortmatrHconv(y) \defeq \matrH(x)|_{x = y}$ satisfies
condition~\eqref{eq:proof:th:girth:4:assumption:y:1}. With this, the
application of convolutional code part of Theorem~\ref{th:girth:4}, along with
Tanner's inequality~\eqref{eq:Tanner:dmin:dfree:1}, yields the QC code part of
Theorem~\ref{th:girth:4}.

\section{Proof of
                Lemma~\ref{lemma:6:cycle:influence:on:permanent:1}}
\label{sec:proof:lemma:6:cycle:influence:on:permanent:1}

We prove only the QC code part of the lemma. The convolutional code part
follows by a similar argument.

We consider only the case where all the entries of $\matr{B}(x)$ are
monomials, \ie,
\begin{align*}
  \matr{B}(x)
    &= \begin{bmatrix}
         x^a & x^b & x^c \\
         x^d & x^e & x^f \\
         x^g & x^h & x^i
       \end{bmatrix}
\end{align*}
for some $a, b, c, d, e, f, g, h, i \in \Zr$. (The discussion for matrices
$\matr{B}(x)$ where some entries are the zero polynomial is analogous.) By
expanding the permanent $\perm\bigl( \matr{B} \bigr)$ of $\matr{B}(x)$ we
obtain
\begin{align*}
  x^{a + e + i} {+}
  x^{b + f + g} {+} 
  x^{c + d + h} {+}
  x^{c + e + g} {+}
  x^{a + f + h} {+}
  x^{b + d + i}.
\end{align*}
Therefore,
\begin{align*}
  \wt\Big( \perm\big( \matr{B}(x) \big) \Big)
    &< \perm\Big( \wt\big( \matr{B}(x) \big) \Big)
\end{align*}
holds if and only if
\begin{align*}
  x^{a + e + i} {+} 
  x^{b + f + g} {+} 
  x^{c + d + h} {+} 
     x^{c + e + g} {+} 
     x^{a + f + h} {+}
     x^{b + d + i}
  < 6.
\end{align*}
For this to hold, there must be at least two monomials that are the same
(in $\shortFtwoxmodr$). Two different cases can happen.
\begin{itemize}

\item Suppose that two monomials like $x^{a + e + i}$ and $x^{b + f + g}$ are
  the same (in $\shortFtwoxmodr$).\footnote{Here, $x^{a + e + i}$ and $x^{b +
      f + g}$ are such that the variables that appear in the exponents are all
    distinct.} Then
  \begin{align*}
    a + e + i
      &= b + f + g \quad \text{(in $\Zr$)},
  \end{align*}
  \ie,
  \begin{align*}
    a - g + i - f + e - b
      &= 0 \quad \text{(in $\Zr$)}.
  \end{align*}
  According to the conditions in~\cite{Tanner:Sridhara:Fuja:01:1,
    Fossorier:04:1}, this is equivalent to the existence of a $6$-cycle in the
  Tanner graph.

\item Suppose that two monomials like $x^{a + e + i}$ and $x^{a + f + h}$ are
  the same (in $\shortFtwoxmodr$).\footnote{Here, $x^{a + e + i}$ and $x^{a +
      f + h}$ are such that there is exactly one variable, \ie, $a$, that
    appears in both exponents.} Then
  \begin{align*}
    a + e + i
      &= a + f + h \quad \text{(in $\Zr$)},
  \end{align*}
  \ie,
  \begin{align*}
    e - h + i - f
      &= 0 \quad \text{(in $\Zr$)}.
  \end{align*}
  According to the conditions in~\cite{Tanner:Sridhara:Fuja:01:1,
    Fossorier:04:1}, this is equivalent to the existence of a $4$-cycle in the
  Tanner graph.

\end{itemize}

The proof is concluded by noting that a $6$-cycle can only appear in a $3
\times 3$ sub-matrix of a type-$1$ polynomial parity-check matrix.

\section{Proof of Theorem~\ref{th:girth:6}}
\label{sec:proof:th:girth:6}

The proof is very similar to the proof of Theorem~\ref{th:girth:4} in
Appendix~\ref{sec:proof:th:girth:4} and so we will only discuss the main steps
of the argument. The following steps are necessary to adapt the proofs of
Theorem~\ref{th:girth:4} in Appendix~\ref{sec:proof:th:girth:4} to the present
corollary.

Convolutional code part:
\begin{itemize}

\item $S_{\nsupseteq \{ 0, 1 \}}$ is replaced by a similarly defined set
  $S_{\nsupseteq \{ 0, 1, 2 \}}$.

\item $S_{\supseteq \{ 0, 1 \}}$ is replaced by a similarly defined
  set $S_{\supseteq \{ 0, 1, 2 \}}$.

\item The line of argument leading to~\eqref{eq:proof:th:girth:4:case:2:3} is
  replaced by the observation that for any $\set{S}' \in S_{\supseteq \{ 0, 1,
    2 \}}$ and any $i \in \set{S}' \setminus \{ 0, 1, 2 \}$ it holds that
  \begin{align*}
    \wt
      \Big(
        \perm\big( \matrH_{\set{S}' \setminus i}(x) \big)
      \Big)
      &\leq 
         \vertsize !
         \, - \, 
         2 (\vertsize {-}3)!.
  \end{align*}
  Therefore, for any $\set{S}' \in S_{\supseteq \{ 0, 1, 2 \}}$ we have
  \begin{align*}
    &\!\!\!\!\!\!\!\!\!\!\!\!\!\!\!\!\!\!\!\!\!
    \sum_{i \in \set{S}'}
      \wt
        \Big(
          \perm\big( \matrH_{\set{S}' \setminus i}(x) \big)
        \Big) \\
      &= \sum_{i \in \{ 0, 1, 2 \}}
           \wt
             \Big(
               \perm\big( \matrH_{\set{S}' \setminus i}(x) \big)
             \Big) 
         \ + \\
      &\quad\ 
         \sum_{i \in \set{S}' \setminus \{ 0, 1, 2 \}}
           \wt
             \Big(
               \perm\big( \matrH_{\set{S}' \setminus i}(x) \big)
             \Big) \\
      &\leq
         3 \vertsize !
         +
         (\vertsize {-}2) (\vertsize ! - 2 (\vertsize {-}3)!) \\
      &\leq 
         (\vertsize {+}1)!
         \, - \, 
         2 (\vertsize {-}2)!.
  \end{align*}

\item The line of argument leading to~\eqref{eq:proof:th:girth:4:case:2:4} is
  replaced by the observation that
  \begin{align*}
    \dfree(\codeCconv)
      &\leq
         \dfree(\codeCconv')
       = \dfree(\codeCconv'') \\
      &\leq
         \big(
           (\vertsize {-}1)+1
         \big)!
       = \vertsize !.
  \end{align*}
  Without loss of generality we can assume that $\vertsize \geq 3$ (otherwise
  a type-$1$ polynomial parity-check matrix $\matrH(y)$ cannot have a
  six-cycle), and so we can upper bound the previous result as follows
  \begin{align*}
    \dfree(\codeCconv)
      &\leq
         (\vertsize {+}1)!
         \, - \, 
         2 (\vertsize {-}2)!.
  \end{align*}
  
\end{itemize}

QC code part:
\begin{itemize}

\item Without loss of generality, we can assume that the rows and columns of
  $\matrH(x)$ are labeled such that
  \begin{align*}
    h_{00}(x)
    h_{11}(x)
    h_{22}(x)
    +
    h_{01}(x)
    h_{12}(x)
    h_{20}(x)
      &= 0
  \end{align*}
  (in $\Ftwoxmodr$). Because multiplying a row of $\matrH(x)$ by an invertible
  element of $\Ftwoxmodr$ produces a polynomial parity-check matrix for the
  same code, and because multiplying a column of $\matrH(x)$ by a monomial
  produces a parity-check matrix of an equivalent code, we can, without loss
  of generality, assume that $\matrH(x)$ is such that
  \begin{align*}
    \begin{bmatrix}
      h_{00}(x) \!\!\! & h_{01}(x) \!\!\! & h_{02}(x) \\
      h_{10}(x) \!\!\! & h_{11}(x) \!\!\! & h_{02}(x) \\
      h_{20}(x) \!\!\! & h_{21}(x) \!\!\! & h_{22}(x) \\
    \end{bmatrix}
      &= \begin{bmatrix}
           1        \!\!\! & 1 \!\!\! & h_{02}(x) \\
           h_{10}(x) \!\!\! & 1 \!\!\! & 1 \\
           1         \!\!\! & h_{21}(x) \!\!\! & 1
    \end{bmatrix}.
  \end{align*}
  (See the end of Appendix~\ref{sec:proof:th:girth:4} for a similar
  reasoning.)  For such a $\matrH(x)$, the polynomial parity-check matrix
  $\shortmatrHconv(y) \defeq \matrH(x)|_{x = y}$ satisfies
  \begin{align*}
    h_{00}(y)
    h_{11}(y)
    h_{22}(y)
    +
    h_{01}(y)
    h_{12}(y)
    h_{20}(y)
      &= 0
  \end{align*}
  (in $\Ftwoylauser$).

\end{itemize}

\section{Proof of Theorem~\ref{th:girth:gen:1}}
\label{sec:proof:th:girth:gen:1}

We prove only the QC code part of the theorem. The convolutional code part
follows by a similar argument.

The proof of the first part of the QC code part of the theorem is very similar
to the proof of Theorem~\ref{th:girth:4} in
Appendix~\ref{sec:proof:th:girth:4} and the proof of Theorem~\ref{th:girth:6}
in Appendix~\ref{sec:proof:th:girth:6}, and is therefore omitted.

So, let us focus on the second part of the QC code part of the
corollary. Because the products on the left-hand and right-hand side
of~\eqref{eq:elementaryproducts:1} are assumed to be nonzero, for every $j \in
\set{R}$ there exists an integer $p_{j,\sigma(j)} \in \Zr$ such that
\begin{align*}
  h_{j,\sigma(j)}(x)
    &= x^{p_{j,\sigma(j)}}.
\end{align*}
Similarly, for every $j \in \set{R}$ there exists an integer $p_{j,\tau(j)}
\in \Zr$ such that
\begin{align*}
  h_{j,\tau(j)}(x)
     = x^{p_{j,\tau(j)}}.
\end{align*}
The condition~\eqref{eq:elementaryproducts:1} can then be rewritten to read
\begin{align*}
  \prod_{j \in \set{R}}
    x^{p_{j,\sigma(j)}}
    &= \prod_{j \in \set{R}}
         x^{p_{j,\tau(j)}}  \quad \text{(in $\shortFtwoxmodr$)}.
\end{align*}
Clearly, this holds if and only if
\begin{align}
  \sum_{j \in \set{R}}
    p_{j,\sigma(j)}
    &= \sum_{j \in \set{R}}
         p_{j,\tau(j)} \quad \text{(in $\Zr$)}.
           \label{eq:elementaryproducts:1:rewriting:1}
\end{align}
Now we define $\pi$ to be the permutation mapping $\pi \defeq \sigma^{-1}
\circ \tau$ from $\set{R}$ to $\set{R}$. (By assumption, $\pi$ is cyclic of
order $R$.) Moreover, we let $j'_0$ be some element of $\set{R}$ and we set
$j'_t \defeq \pi^t(j'_0)$, $t = 1, \ldots, R{-}1$. Then, the condition
in~\eqref{eq:elementaryproducts:1:rewriting:1} holds if and only if
\begin{align*}
  \sum_{t = 0}^{R-1}
    p_{j'_t, \sigma(j'_t)}
    &= \sum_{t = 0}^{R-1}
         p_{j'_t, \tau(j'_t)}
           \quad \text{(in $\Zr$)},
\end{align*}
which holds if and only if
\begin{align}
  \sum_{t = 0}^{R-1}
    \left(
      p_{j'_t, \sigma(j'_t)}
      -
      p_{j'_t, \tau(j'_t)}
    \right)
    &= 0 \quad \text{(in $\Zr$)}.
         \label{eq:elementaryproducts:1:rewriting:2}
\end{align}
Because we assumed that $\pi = \sigma^{-1} \circ \tau$ is a cyclic permutation
of order $R$, the condition in~\eqref{eq:elementaryproducts:1:rewriting:2} is
equivalent to Tanner's condition on the existence of a $2 R$-cycle,
see~\cite{Tanner:Sridhara:Fuja:01:1, Fossorier:04:1}. (Note that
$\sigma(j'_{t+1}) = \sigma \bigl( \pi(j'_{t}) \bigr) = \tau(j'_{t})$ and that
$\sigma(j'_{0}) = \sigma \bigl( \pi(j'_{R-1}) \bigr) = \tau(j'_{R-1})$.)

\section{Graph Covers}
\label{sec:graph:covers:1}

This appendix collects some results that are used in
Section~\ref{sec:type:1:qc:ldpc:codes:based:on:double:cover:1}. The focus is
on QC codes, however, similar results can also be stated for convolutional
codes.

Let $\code{C}$ be a QC code with polynomial parity-check matrix
\begin{align*}
  \matrH(x)
    = \big[
        h_{j,i}(x)
      \big]_{j,i}
      \in \shortFtwoxmodr^{\vertsize  \times \horsize}.
\end{align*}
We define the decomposition
\begin{align*}
  \matrH(x)
    &= \matrH^{(1)}(x) + \matrH^{(2)}(x)
       \ \ \left(\text{in $\shortFtwoxmodr^{\vertsize  \times \horsize}$}
         \right)
\end{align*}
with matrices
\begin{align*}
  \matrH^{(1)}(x)
    &= \left[
         h^{(1)}_{j,i}(x)
       \right]_{j,i}
       \in \shortFtwoxmodr^{\vertsize  \times \horsize}
\end{align*}
and
\begin{align*}
  \matrH^{(2)}(x)
    &= \left[
         h^{(2)}_{j,i}(x)
       \right]_{j,i}
       \in \shortFtwoxmodr^{\vertsize  \times \horsize}.
\end{align*}
Based on this decomposition, we define a new code $\code{\tilde C}$ with the
polynomial parity-check matrix
\begin{align*}
  \matr{\tilde H}(x)
    &\defeq
       \begin{bmatrix}
         \matrH^{(1)}(x) & \matrH^{(2)}(x) \\
         \matrH^{(2)}(x) & \matrH^{(1)}(x)
       \end{bmatrix}
       \in \shortFtwoxmodr^{2\vertsize  \times 2\horsize}.
\end{align*}

\begin{lemma}
  \label{lemma:double:cover:minimum:distance:1}

  The minimum Hamming distances of $\code{C}$ and $\code{\tilde C}$ satisfy
  \begin{align}
    \dmins(\code{C})
      &\leq \dmins(\code{\tilde C})
       \leq 2 \cdot \dmins(\code{C}).
         \label{eq:double:cover:minimum:distance:relationship:1}
  \end{align}
\end{lemma}

\begin{IEEEproof}
  Let us start by proving the first inequality
  in~\eqref{eq:double:cover:minimum:distance:relationship:1}. Let $\vtc(x) =
  (\vc^{(1)}(x), \vc^{(2)}(x))$ be a codeword in $\code{\tilde C}$ with
  Hamming weight $\wH\bigl( \vtc(x) \bigr) = \wH\bigl( \vc^{(1)}(x) \bigr) +
  \wH\bigl( \vc^{(2)}(x) \bigr) = \dmins(\code{\tilde C})$. We show that
  $\vc(x) \defeq \vc^{(1)}(x) + \vc^{(2)}(x) \in \code{C}$. Indeed, because
  $\vtc(x) \in \code{\tilde C}$, we have (in $\shortFtwoxmodr$)
  \begin{align}
    \matrH^{(1)}(x) \cdot \vc^{(1)}(x)^\tr
    +
    \matrH^{(2)}(x) \cdot \vc^{(2)}(x)^\tr
      &= \vect{0}^\tr, 
           \label{eq:double:cover:parity:check:equation:1} \\
    \matrH^{(2)}(x) \cdot \vc^{(1)}(x)^\tr
    +
    \matrH^{(1)}(x) \cdot \vc^{(2)}(x)^\tr
      &= \vect{0}^\tr.
           \label{eq:double:cover:parity:check:equation:2}
  \end{align}
  Adding these two equations we obtain (in $\shortFtwoxmodr$)
  \begin{align*}
    \left(
      \matrH^{(1)}(x)
      +
      \matrH^{(2)}(x)
    \right)
    \cdot
    \left(
      \vc^{(1)}(x)
      +
      \vc^{(2)}(x)
    \right)^\tr
      &= \vect{0}^\tr,
  \end{align*}
  showing that $\vc(x) \in \code{C}$. If $\vc(x) \neq \vect{0}$ then
  \begin{align*}
    \dmins(\code{C})
      &\leq
         \wH\big( \vc(x) \big) \\
      &= \wH\left( \vc^{(1)}(x) +  \vc^{(2)}(x) \right) \\
      &\leq
         \wH\left( \vc^{(1)}(x) \right) +  \wH\left( \vc^{(2)}(x) \right) \\
      &= \wH\big( \vtc(x) \big)
       = \dmins(\code{\tilde C}),
  \end{align*}
  thus proving the first inequality
  in~\eqref{eq:double:cover:minimum:distance:relationship:1}. If $\vc(x) =
  \vect{0}$ then $\vc^{(1)}(x) = \vc^{(2)}(x)$ and so
  both~\eqref{eq:double:cover:parity:check:equation:1}
  and~\eqref{eq:double:cover:parity:check:equation:2} can be rewritten to read
  (in $\shortFtwoxmodr$)
  \begin{align*}
    \left(
      \matrH^{(1)}(x)
      +
      \matrH^{(2)}(x)
    \right)
    \cdot
      \vc^{(1)}(x)^\tr
      &= \vect{0}^\tr,
  \end{align*}
  showing that $\vc^{(1)}(x) = \vc^{(2)}(x) \in \code{C}$. With this, 
  \begin{align*}
    \dmins(\code{C})
       \leq
         \wH\left( \vc^{(1)}(x) \right)
      &< 2 \cdot \wH\left( \vc^{(1)}(x) \right) \\
      &= \wH\big( \vtc(x) \big)
       = \dmins(\code{\tilde C}),
  \end{align*}
  thus proving the first inequality
  in~\eqref{eq:double:cover:minimum:distance:relationship:1}.

  We now prove the second inequality
  in~\eqref{eq:double:cover:minimum:distance:relationship:1}. Let $\vc(x)$ be
  a codeword in $\code{C}$ with Hamming weight $\wH\bigl( \vc(x) \bigr) =
  \dmins(\code{C})$ and define $\vtc(x) \defeq \bigl( \vc(x), \vc(x) \bigr)$.
  We show that $\vtc(x) \in \code{\tilde C}$. Indeed, because $\vc(x) \in
  \code{C}$, we have (in $\shortFtwoxmodr$)
  \begin{align*}
    \matrH(x) \cdot \vc(x)^\tr
      &= \vect{0}^\tr. 
  \end{align*}
  Therefore  (in $\shortFtwoxmodr$),
  \begin{align*}
    \left(
      \matrH^{(1)}(x) + \matrH^{(2)}(x)
    \right)
      \cdot
      \vc(x)^\tr
      &= \vect{0}^\tr,
  \end{align*}
  and so  (in $\shortFtwoxmodr$)
  \begin{align*}
    \matrH^{(1)}(x) \cdot \vc(x)^\tr
    +
    \matrH^{(2)}(x) \cdot \vc(x)^\tr
      &= \vect{0}^\tr, \\
    \matrH^{(2)}(x) \cdot \vc(x)^\tr
    +
    \matrH^{(1)}(x) \cdot \vc(x)^\tr
      &= \vect{0}^\tr. 
  \end{align*}
  which are exactly the equations that $\vtc(x)$ must satisfy in order to be a
  codeword in $\code{\tilde C}$. Therefore, because $\vtc(x) \neq \vect{0}$,
  \begin{align*}
    \dmins(\code{\tilde C})
      &\leq
         \wH\big( \vtc(x) \big)
       = 2 \cdot \wH\big( \vc(x) \big)
       = 2 \cdot \dmins(\code{C}),
  \end{align*}
  thus proving the second inequality
  in~\eqref{eq:double:cover:minimum:distance:relationship:1}. 
\end{IEEEproof}

\mbox{}

Assume that the matrices $\matrH^{(1)}(x)$ and $\matrH^{(2)}(x)$ are such that
when $h^{(1)}_{j,i}(x)$ and $h^{(2)}_{j,i}(x)$ are added to obtain
$h_{j,i}(x)$, then no terms cancel for any $j$ and and $i$. In this case it
can easily be seen that the Tanner graph of $\matr{\tilde H}(x)$ is a double
cover of the Tanner graph of $\matrH(x)$. This means that Lemma
\ref{lemma:double:cover:minimum:distance:1} relates the minimum Hamming
distance of a Tanner graph and the minimum Hamming distance of a certain
double cover of that Tanner graph.

There is another way to obtain the same double cover (up to relabeling of the
coordinates). Namely, based on code $\code{C}$ we define the code $\code{\tilde
  C'}$ with the polynomial parity-check matrix
\begin{align*}
  \matr{\tilde H}'(x)
    &\in \shortFtwoxmodr^{2\vertsize  \times 2\horsize}.
\end{align*}
This time the matrix $\matr{\tilde H}'(x)$ is obtained from $\matrH(x)$ by
replacing, for each $j$ and each $i$, the $1 \times 1$ entry
\begin{align*}
  h_{j,i}(x)
    &= h^{(1)}_{j,i}(x)
       +
       h^{(2)}_{j,i}(x)
\end{align*}
by the $2 \times 2$ entry 
\begin{align*}
  \begin{pmatrix}
    h^{(1)}_{j,i}(x) & h^{(2)}_{j,i}(x) \\
    h^{(2)}_{j,i}(x) & h^{(1)}_{j,i}(x)
  \end{pmatrix}.
\end{align*}
It can easily be checked that $\matr{\tilde H}'(x)$ equals $\matr{\tilde
  H}(x)$ up to reshuffling of rows and columns. Therefore, the Tanner graphs
of $\matr{\tilde H}(x)$ and of $\matr{\tilde H}'(x)$ are isomorphic, showing
that they define the same double cover of the Tanner graph of $\matrH(x)$ (up
to relabeling of the bit and check nodes).

Let us remark that~\cite{Tai:Lan:Zeng:Lin:Abdel-Ghaffar:06,
  Ivkovic:Chilappagari:Vasic:08:1, Asvadi:Banihashemi:AhmadianAttari:11:1}
consider similar $2$-covers.\footnote{The
  paper~\cite{Asvadi:Banihashemi:AhmadianAttari:11:1} also considers
  higher-degree covers, in particular covers whose degree is a power of $2$.}
Namely, for some (scalar) parity-check matrix $\matrH$, these authors first
constructed the trivial $2$-cover
\begin{align*}
  \matr{\tilde H}
    &= \begin{bmatrix}
         \matrH & \matzero \\
         \matzero & \matrH
       \end{bmatrix}.
\end{align*}
Then, in a second step they split $\matrH$ into $\matrH = \matrH' + \matrH''$,
where the matrices $\matrH'$ and $\matrH''$ were chosen such that the nonzero
entries do not overlap. Finally, they formulated a modified $2$-cover as
follows
\begin{align*}
  \matr{\tilde H}'
    &= \begin{bmatrix}
         \matrH'  & \matrH'' \\
         \matrH'' & \matrH'
       \end{bmatrix}.
\end{align*}

We conclude this appendix by remarking that similar results can be proved for
general $M$-covers, where $M$ is a power of $2$, by iterating the results of
this section multiple times.

\ifCLASSOPTIONcaptionsoff
  \newpage
\fi

\end{document}